# Edge Factors: Scientific Frontier Positions of Nations

Mikko Packalen
University of Waterloo

17 June 2018

**Abstract**

A key decision in scientific work is whether to build on novel or well-established ideas. Because exploiting new ideas is often harder than more conventional science, novel work can be especially dependent on interactions with colleagues, the training environment, and ready access to potential collaborators. Location may thus influence the tendency to pursue work that is close to the edge of the scientific frontier in the sense that it builds on recent ideas. We calculate for each nation its position relative to the edge of the scientific frontier by measuring its propensity to build on relatively new ideas in biomedical research. Text analysis of 20+ million publications shows that the United States and South Korea have the highest tendencies for novel science. China has become a leader in favoring newer ideas when working with basic science ideas and research tools, but is still slow to adopt new clinical ideas. Many locations remain far behind the leaders in terms of their tendency to work with novel ideas, indicating that the world is far from flat in this regard.

**Acknowledgements:** I thank Jay Bhattacharya, Bruce Weinberg, Partha Bhattacharyya, Richard Freeman, Horatiu Rus, Joel Blit, David Autor, Larry Smith and Peter Tu for discussions. I acknowledge financial support from the National Institute on Aging grant P01-AG039347.

## 1. Introduction

Knowledge production is an increasingly global endeavor. In spite of robust increases in scientific production by the traditional leaders – including the United States, the United Kingdom, and Japan – their relative share has decreased in recent decades because the pace of growth in science by other nations – including China, South Korea, India, and Brazil – has been even more rapid (Freeman, 2013; National Science Board, 2016). The share of international collaborations has also increased, as has the share of citations to papers with foreign authors (Freeman, 2013, National Science Board, 2016). This spread of knowledge production has not been unexpected. It was anticipated long ago by Marshall (1920) that improved communication technologies would make it easier to learn about new discoveries regardless of location and that this would lead to the pursuit of creative work in more diverse places.

While this perspective suggests a diminishing influence for location in scientific work, location may in fact continue to have considerable import in science. This is because learning about which new ideas exist may not have been an important benefit of location for quite some time and because location likely still impacts the fertility of creative work in other important ways. Lucas (2004), for example, has emphasized the continuing dependence of knowledge production activities on location through the benefits that accrue from daily interactions with colleagues, the training environment, and ready access collaborators.

One potential remaining influence of location stems from the fact that when new ideas are first discovered, they are often raw and poorly understood. The ideas only gradually mature into useful advances after a community of scientists tries them out and develops them. But such work is hard, harder than work that builds on well-established ideas. One indication that work that tries out new ideas is indeed harder than more conventional science is that such work is linked with larger team size (Packalen and Bhattacharya, 2016). Thus, when a scientist seeks to build on a recent advance, it is beneficial to be surrounded by a community of scholars with whom to debate about which new ideas to try out and how (Marshall, 1920; Kuhn, 1962, 1977; Usher, 1929). Daily interactions with colleagues, the training environment, and ready access to potential collaborators thus become especially important in work that is close to the edge of the scientific frontier in the sense that the work builds on recent advances.

Because such local factors influence the fertility of the debates that seek to unlock the mysteries of new ideas, the tendency to work with new ideas can be expected to vary by location. This mechanism – and thus the import of location – may even be increasingly influential, for Jones (2010) shows that reaching the scientific frontier now involves even more work than before as evidenced by increases in training times, specialization, and teamwork.

Therefore, even as the pursuit of science spreads to more diverse places, location may well continue to have an important influence on *what kind of science* is pursued – through the impact that location may have on the ability to work with novel ideas. Identifying where barriers to knowledge adoption still exist is thus crucial for understanding the role of location in knowledge production and for designing policies that can help eliminate the remaining barriers.

We calculate each nation's propensity to publish biomedical work that is close to the edge of the scientific frontier in the sense that it builds on relatively recent ideas. The results reveal each nation's position on the scientific frontier: what share of its contributions to biomedical science build on relatively new ideas vs. well-established ideas. Our empirical analysis is focused on biomedicine because it is an important area of science and because of the availability of the Pubmed/MEDLINE database on over 24 million biomedical research papers.

We refer to our constructed measure of novelty as *the edge factor*. Whereas the familiar impact factor measures scientific influence (Garfield, 1955, 1972), the edge factor measures an aspect of novelty of scientific work – the tendency to build on ideas close to the edge of the scientific frontier. A common characteristic for these two measures is that for each entity they both quantify the average of a characteristic (rather than the total number of novel contributions or the total number of received citations).

We selected countries as the unit of analysis because borders continue to influence scientist interactions and because many important science policy decisions are set at the national level. However, similar to the impact factor, the edge factor too can be constructed also for many other units of analyses. For example, the approach can be utilized to evaluate the novelty of research published by a journal or institution. It can also be applied to analyze the novelty of individual scientists' publications and examine which scientist-level characteristics promote the trying out of new ideas. Furthermore, the edge factor can be utilized to compare to what extent funding agencies succeed in their often-stated aim of supporting novel work. The focus on countries in the present paper is thus just one possible application. But this focus is useful because it provides way to utilize this concept – the edge factor – to shed light on a long-standing question about to what extent the spread of modern communication technologies has eradicated location-based barriers to the adoption of new ideas. At the same time, this application illustrates the potential of the edge factor in other contexts.

We emphasize that the impact factor and the edge factor capture distinct aspects of science – impact and novelty – regardless of the specific setting in which they are employed. Moreover, the edge factor and the impact factor are complementary tools in policy evaluation and design. For optimal science policy requires that both influence and novelty are rewarded. One reason why rewarding influence alone is not enough is that rewarding novelty directly helps solve a coordination problem that is inherent in the formation of a vibrant scientific community to a new area of investigation (Besancenot and Vranceaunu, 2015; Packalen and Bhattacharya, 2017). A singular focus on citation counts can lead to stagnant science because impact factors under-reward scientists who try out new ideas, thereby stifling work that helps ideas mature and makes more meaningful advances possible. Another related reason to reward novel work is that useful work that tries out a new idea need not be influential in the traditional sense; such work can have scientific value – in terms of helping unlock the mysteries of the new idea – even when it merely demonstrates which research paths do not work.

In recent decades, many – including the editor of *Science* (Alberts, 2013) – have raised alarm about the science community's obsession with impact factors. The obsession with impact may have already led to less healthy science, as the rise of citation metrics has coincided with a decline in the novelty of biomedicine (Rzhetzky et al., 2015). By using measures like the edge factor in conjunction with impact-based metrics, university administrators and funding agencies can strike a better balance between rewarding innovative but risky work that develops ideas early on and rewarding work that takes advantage of the ideas in their more mature stages. Making reward structures even slightly more favorable to scientific novelty would encourage scientists to pursue more innovative research paths and lead to healthier, less stagnant, science. To be sure, the edge factor is not meant to displace the impact factor. Instead, it is ideally used as a complementary metric that captures a different aspect of science.

As in the closest prior work (Packalen and Bhattacharya, 2017), the focus here is on the novelty of idea inputs, as opposed to the novelty of the combination of idea inputs. Novelty of combinations is a focus in several recent analyses (Wang et al., 2016; Lee et al., 2015; Rzhetzky et al., 2015; Boudreau et al., 2016; Foster et al., 2015). Both foci come with their advantages, as discussed in Packalen and Bhattacharya (2017). The focus on the use of new ideas makes it possible to include on a larger number of ideas in the analysis than is computationally feasible in an analysis of combinatorial novelty. Analysis of the use of new ideas is also important because the trying out of new ideas is so central to scientific progress. For without new ideas science eventually stagnates – combinatorial novelty alone cannot overcome it.

Similar to the closest prior work (Packalen and Bhattacharya, 2017)., we use text analysis to determine the ideas that each paper built upon and also the vintage of those ideas. There are two main differences between the present study and this prior work. First, there is a shift in substantive focus – from ranking journals to ranking nations. Second, in the present approach the novelty of each contribution is allowed to depend not just on the vintage of ideas employed in it but also on what types of ideas they are. This is important because a paper that employs a 10-year old research tool may represent novel work but the same need not be true for a paper that examines a gene of the same vintage. This methodological innovation yields a more finely grained measurement of each entity's distance to the edge of the scientific frontier. Importantly, this methodological advance can be utilized in various applications beyond the present analysis.

**2. Methods**

In this section, we first describe the data sources (MEDLINE biomedical publications database, NLM Journal Categories, the UMLS Metathesaurus, and the MeSH vocabulary). In subsection 2.2 we explain why the analysis is focused on years 1988-2016, and in subsection 2.3 we explain why the location of each contribution is determined based on the first author's affiliation. In subsection 2.4 we explain how the research area of each contribution is determined.

In subsection 2.5 we explain how we use the UMLS metathesaurus to determine the ideas that each paper built upon and also how the vintage of each idea is determined. In subsection 2.6 we explain how we define a contribution (by determining links from each paper to each research area and each idea category), how we measure the idea category of each idea based on the UMLS metathesaurus, why the resulting contribution-level analysis is preferable to a paper-level analysis, and also how the novelty of each contribution is determined.

In subsection 2.7 we explain how the overall edge factor for each nation is calculated: we first determine the nation's edge factor separately for each (idea category, research area) pair based on all the nation's contributions linked to that pair; afterwards we calculate the nation's overall edge factor as a weighted sum of its edge factors across all (idea category, research area) pairs. Finally, in subsection 2.8 we explain how the approach developed here differs from the approach used in closest prior work.

### 2.1 Data Sources

#### 2.1.1 MEDLINE Database on Biomedical Research Publications

Our source for information on scientific publications is the MEDLINE database (https://www.nlm.nih.gov/bsd/pmresources.html). MEDLINE is a comprehensive database on life sciences with a focus on biomedicine in particular. The database contains information on over 24 million journal articles.

For each journal article in MEDLINE, we make use the following variables: publication year, affiliation of the first author, text of the title and abstract, journal where the article was published, and MeSH keywords. The acronym "MeSH" stands for *Medical Subject Headings*; the MeSH vocabulary is a controlled vocabulary of over 87,000 terms (https://www.nlm.nih.gov/mesh/). Publications in MEDLINE indexed with MeSH keywords; we use the MeSH keywords to determine article type (we focus the analysis on original research articles) and whether an article represents applied or basic science (section 2.4).

#### 2.1.2 Broad Subject Terms for MEDLINE Journals

Our source for the research area of each article is the broad subject terms that are assigned by the National Library of Medicine for journals in the MEDLINE database (https://wwwcf.nlm.nih.gov/serials/journals/index.cfm). We show further below how articles in the MEDLINE database are distributed across the journal categories in this database (section 2.4).

### 2.1.3 Unified Medical Language System (UMLS) Metathesaurus

As our source for information on which words and word sequences represent meaningful concepts in biomedicine and which concepts are synonyms, we use the 2017 version of the Unified Medical Language System (UMLS) metathesaurus (https://www.nlm.nih.gov/research/umls/). The UMLS metathesaurus links over 5 million terms that appear in one or more of over 150 medical vocabularies.

In addition to determining the synonyms for each term, the UMLS database assigns each term to one or more of 127 semantic types (https://semanticnetwork.nlm.nih.gov). We use the semantic type of each term to represent the idea category of the term (section 2.6.1). Further below we list examples of ideas and idea categories captured by this approach (section 2.6.2).

## 2.2 Sample of Papers

When we determine the vintage of each idea (section 2.5.2), we use the sample of all papers in the MEDLINE database. By contrast, when we calculate for each location its propensity to publish novel work, we limit the sample of papers in several ways. First, we limit the analysis to original research papers, thereby excluding editorials, reviews, etc. However, in a robustness analysis, we include all papers in the sample. Second, we limit the analysis to papers published during 1988-2016. This is because the coverage for affiliation data in the MEDLINE database begins in 1988. Third, we limit the analysis to papers for which the available text on the title and the abstract of the article in the database includes at least 200 characters and no more than 5000 characters. However, in a robustness analysis, we conduct the analysis without this character limit. The number of articles that are included in our main specification is shown by publication year in Figure S1 (Web Appendix).

## 2.3 Country of Each Scientific Publication

We assign each paper to a country based on the affiliation string for the first author of the paper. We limit the analysis to first authors because for most papers published before 2014 the affiliation information in MEDLINE is limited to the first author of each paper. Figure S2 (Web

Appendix) shows by publication year the share of papers that we were able to match to a country. For ease of exposition we limit the number of locations by combining some countries that publish a smaller number of biomedical publications to regions. Figure S3 (Web Appendix) shows the share of papers by location (country or region) and time period.

### 2.4 Journal Categories and Journal Category Groups

We use the journal categories (Broad Subject Terms) to represent the research area of each paper. On average, each original research article published during 2015-2016 is linked to 1.49 journal categories. Table S1 (Web Appendix) shows the distribution of links from papers to journal categories during this time period. As is discussed in detail further below (section 2.6.1), papers from journals that are linked to multiple journal categories are considered to have contributed to multiple research areas.

In our main specification, all journal categories are included in the analysis. In secondary analyses, we conduct three separate analyses – each limits the analysis to one of the following three groups of journal categories: "Applied", "Basic Science", and "Other (Both Applied and Basic Science)".

To conduct these secondary analyses, we assign each journal category to one of the three journal category groups. Here we make use of the MeSH keywords affixed to each MEDLINE article and the "A-C-H" model (Weber, 2014) that classifies papers along the translational axis based on the MeSH keywords. Specifically, using the MeSH codes we first determine each paper's position on the translational axis as specified by the A-C-H model: "H status" (human) is assigned to papers with either the MeSH code B01.050.150.900.649.801. 400.112.400.400 (Human) or the Mesh code M01 (Person), "C status" (cells and molecules) is assigned to papers with any of the following MeSH codes (or codes that appear in the MeSH subtrees of these MesH codes): A11 (Cells), B02 (Archaea), B03 (Bacteria), B04 (Viruses), G02.111.570 (Molecular Structures), and G02.149 (Chemical Processes), and "A status" (animal) is assigned to papers with the MeSH code B01 (Eukaryota) and papers with any of the codes in the subtree of this MeSH code B01 except the aforementioned MeSH code for "Human".

We thus construct three separate indicator variables ("H status", "A status", "C status"). In the A-C-H model, papers with "H status" have an applied aspect to them, and papers with either “A status” or “C status” have a basic science aspect to them. More than one of these indicator variables will be positive for papers that have both an applied and a basic science aspect to them.

For each journal category, we next calculate the average of each of these three dummy variables (“H status”, “A status”, “C status”) among all papers linked to that journal category. Denoting these variables as “Average H status”, “Average A status”, and “Average C status”, we use them to classify journal categories to three journal category groups as follows. Journal categories that satisfy conditions “Average H status > Average C status” and “Average H status > 0.2” are assigned to journal category group “Applied”. Journal categories that satisfy “Average H status < Average C status” and “Average A status < 0.8” and “Average C status > 0.5” are assigned to journal category group “Basic Science”. (We thus exclude journal categories that focus heavily on veterinary medicine from this category even though such journal categories are located early along the translational axis in the A-C-H model; this happens in the A-C-H model because the model does not distinguish between veterinary medicine and animal studies as pre-cursor to human medicine). The remaining journal categories are assigned to journal group category “Other (Both Applied and Basic Science)”. The result of this approach for determining the journal category group of each journal category is shown in the last column of Table S1.

## 2.5 Identifying Ideas and the Vintage of Ideas

### 2.5.1 Using the UMLS metathesaurus to identify ideas from text

We employ text analysis to discern which ideas each research paper built upon. We treat each of the 5+ million terms in the comprehensive United Medical Language System (UMLS) meta-thesaurus as representing ideas. To identify which of these ideas each research paper in the MEDLINE database built upon, we search the title and abstract of each publication for all the terms in the UMLS metathesaurus.

Thus, the first step in the text analysis is to determine for each article in the MEDLINE database which UMLS terms appear in it. Further below we also show a list of examples of ideas identified by this approach (section 2.6.2).

### 2.5.2 Calculating the vintage of each idea

The vintage of the idea represented by a UMLS term is determined based on how long ago the UMLS term was first mentioned in a biomedical research paper. We interpret the mention of a relatively new term as indicative of work that builds on ideas close to the edge of the scientific frontier. We refer to the year of first appearance of a term as the cohort year of the term. In a robustness analysis, we set the cohort year of each term as the earliest year the UMLS term or any of its synonyms appears in the MEDLINE data (synonyms are determined based on the synonym information in the UMLS metathesaurus). The results from this robustness analysis show that our results are not driven by relabeling of old ideas.

Because of the sparsity of publications in MEDLINE with a publication year before 1946, the cohort year of ideas (i.e. the year of first appearance) does not reflect the ideas' true vintage well for ideas that are new to biomedicine before 1950. Thus, we exclude from the analysis all terms with cohort before 1950.

Further below we show examples of cohort years assigned to terms using this approach (section 2.6.2).

## 2.6 Contribution-Level Analysis

### 2.6.1 Defining a contribution as a link from a paper to an (idea category, research area) pair

In determining the novelty of biomedical work, we seek to control for the idea category of each idea. Thus, we aim to compare the use of novel ideas against the use of more established ideas from the same idea category. The rationale for seeking to control for the idea category is the following: how recent ideas should be considered novel depends on what type of an idea it is. For example, a paper that employs a 10-year old research tool may represent novel work but the same need not be true for a paper that examines a gene of the same vintage.

To control for the idea category in the present analysis, we take advantage of the fact that the UMLS metathesaurus classifies terms to 127 semantic types (these are listed further below in section 2.6.2). We treat each of these UMLS semantic types as representing one idea category.

We make use of these idea categories as follows. After determining which UMLS terms are mentioned in each paper, we determine which UMLS categories are represented by these terms. We then treat a paper that mentions terms from K different idea categories as K separate contributions. The underlying assumption in this approach is that work that mentions at least one idea from an idea category advances our understanding of how ideas from that idea category work. Thus, work that mentions ideas from multiple categories advances our understanding on multiple dimensions; hence we treat it as multiple contributions.

Table S2 (Web Appendix) shows the number of links to each idea category from papers published during 2015-2016. As was mentioned above in section 2.5.2, we only include in the analysis those terms that have cohort year 1950 or later.

In our main analysis, we calculate the overall edge factor based on links to any of the 127 idea categories (the edge factor is an average across all idea categories). In a secondary analysis, we calculate the edge factor separately for each of the following four groupings of idea categories: “Clinical and Anatomy”, “Drugs and Chemicals”, “Basic Science and Research Tools”, and “Miscellaneous”. To conduct this analysis, we link each idea category to one of these four idea category groups. The last column of Table S2 shows which idea category belongs to which idea category group. This secondary analysis reveals for each country whether any barriers to new idea adoption are limited by to certain types of ideas. The decision to link each of the 127 idea categories to one of four groups was made for expositional purposes – it implies that for each country we must report four separate numbers.

What kind of work should be considered novel is likely to depend also on the research area. For example, use of a 10-year old research tool may be novel work in public health research but not in biotechnology research. To address this issue, we also determine the links from papers to research areas. We use the National Library of Medicine (NLM) journal categories as proxies for research areas (these journal categories were listed in section 2.4, Table S1).

Thus, after determining the ideas mentioned in each paper, we determine which idea categories are linked to these ideas as well as which research areas are linked to the journal where the paper is published. We define a *contribution* as an (idea category, research area) pair linked to a paper.

In our approach, a paper is considered to contribute to our understanding of all the (idea category, research area) pairs linked to it. A paper can make multiple contributions, depending on how many (idea category, research area) pairs are linked to it. A paper that mentions ideas from K idea categories, and is published in a journal that is linked to J journal categories, is treated as K*J separate contributions.

Note that a paper that mentions multiple ideas from an idea category results in the same number contributions as a paper that mentions only one idea from the idea category.

The number of links listed in the second column of Table S1 is the number of links to (idea category, research area) pairs associated with each research area. Similarly, the number of links listed in the second column of Table S2 is the number of links to (idea category, research area) pairs associated with each idea category. On average, each paper published during 2015-2016 is linked to linked to 6.26 (idea category, research area) pairs. Therefore, in our approach each paper is, on average, counted as 6.26 contributions.

A limitation of the approach pursued here is the inherent assumption that, synonyms aside, all ideas within an idea category are treated as though they are equally close to one another although in reality some are closer than others. Similarly, all ideas within a category are treated as though they are equally distance is the same to all ideas outside the idea category. A potentially important direction for future work is to explore a more fine-grained approach that either calculates a pairwise between each idea category or calculates a pairwise distance between each idea.

When determining whether a contribution represents novel work, we only consider the age of the newest term linked to the (idea category, research area) pair from the paper in question. Researcher's choice is between using any new ideas or only well-established ideas from this idea category. This is discussed in more detail next.

### 2.6.2 The novelty of a contribution

Above we defined a contribution as a link from a paper to an (idea category, research area) pair; these links are inferred from the UMLS terms that appear in the title and abstract of the paper. The novelty of each contribution is determined in three steps.

Step 1. Age of each UMLS term that links a paper to the (idea category, research area) pair. First, for each contribution associated with a paper, we determine the age of each term that links the paper to the (idea category, research area) pair in question. Age of each term is calculated by subtracting the cohort year of the term from the publication year of the paper.

Step 2. Age of the newest UMLS term that links a paper to the (idea category, research area) pair. Second, for each contribution we determine the age of the newest term that links the paper to the (idea category, research area) pair. We refer to the cohort year of the newest term that links a paper to the (idea category, research area) pair as the cohort year of the contribution.

Step 3. Novelty of the contribution relative to other contributions to the (idea category, research area) pair among papers published in the same year. The relative novelty of a contribution is then determined by comparing the vintage of the contribution to the vintages of all the other contributions linked to the same (idea category, research area) pair, among papers published in the same year. The interpretation is that a paper that links to an idea category reflects a choice faced by a scientist: one can choose work with at least one relatively new idea from this idea category, or one can choose to work with only well-established ideas from this idea category. The comparison is also limited by research area because whether the use of an idea represents novel work is expected to depend on the context where it is used. The reason for limiting the comparison to papers published in the same year is obvious: because the rate of scientific progress need not be the same over time, the use of a 10-year old research tool may represent novel work in one year but not in some other year.

Having determined all contributions linked to an (idea category, research area) pair among papers published in the same year, we order the contributions based on their vintage (age of the newest term linked to that (idea category, research area) pair from each paper). We then construct an indicator variable that captures the relative novelty of each contribution: in our baseline specification, contributions that are in the top 5% based on their vintage are considered novel work (the indicator variable is 1 for such contributions and 0 otherwise). In robustness analyses, we construct the indicator variable using alternative choices for the cutoff percentile (top 1%, top 5%, or top 20%). Figure S4 (Web Appendix) shows the distribution of the cohort of *all* contributions based on papers published during 2015-2016. Figure S5 (Web Appendix) in turn shows the distribution of cohort of *novel* contributions when novelty of a contribution is determined based on the top 5% status and also the when novelty of a contribution is determined based on one of the alternative cutoffs (top 20%, top 10%, or top 1%). When the top 5% cutoff is used, then for all post-2004 cohorts the majority of contributions with that cohort are deemed novel by our approach, and for all pre-2004 cohorts at most a minority of contributions are deemed novel by our approach.

Table S3 (Web Appendix) shows examples of ideas, as represented by UMLS terms, captured by our approach. The table also shows the idea category of each term. Some terms appear multiple times because these terms are linked to multiple UMLS categories by the UMLS metathesaurus. As in related prior work (*15*), the list of terms shows that the approach used here captures ideas that are widely recognized to have been important inputs in biomedical work in recent decades (for expositional reasons the list is focused on popular ideas – there are of course also many unpopular, less important ideas that are captured by our approach) and that for most terms the cohort year assigned to the term reflects the era when the idea represented by the term entered biomedicine.

## 2.7 Calculation of the Edge Factor

### 2.7.1 Novelty of a nation's contributions linked to a specific each (idea category, research area) pair

Normalization of contribution-level novelty indicators. Having determined the contributions of each paper (i.e. which (idea category, research area) pairs are linked to from each paper) and which contributions are novel (i.e. which contributions have the top 5% status based on their vintage), we next normalize the novelty variable within contributions to each (idea category, research area) pair so that the average of the normalized novelty variable is 100 within each (idea category, research area) pair. In implementing the normalization, we combine data from multiple years. For example, in our main specification we combine data from 2015-2016.

Location-level novelty scores for each (idea category, research area) pair. Using the normalized contribution-level novelty variable, we then calculate for each location its propensity for novel work within each (idea category, research area) pair. That is, for each location we calculate the mean of the normalized novelty variable based on all of the location's contributions linked to a specific (idea category, research area) pair. We refer to each such average of the novelty variables as the edge factor of the location for the specific (idea category, research area) pair. An edge factor above (below) 100 indicates an above (below) average tendency for work that builds on relatively novel ideas. In our main specification, these edge factors are calculated based on papers published during 2015-1016.

### 2.7.2 Calculating the overall edge factor

Having determined the relative novelty of each location's contributions separately for each (idea category, research area) pair – the location's edge factor for that (idea category, research area) pair – we construct the overall edge factor for each nation as a weighted sum of these (idea category, research area) pair specific edge scores.

Weights. In our main specification, we use as weights the frequency at which each (idea category, research area) pair is encountered in biomedicine. In other words, the weight of an edge factor for an (idea category, research area) pair is the total number of papers linked to it from any location during the time period.

A justification for selecting these weights is that those (idea category, research area) pairs that are encountered more often in biomedicine are, by revealed preference, considered more important by scientists. The ability to pursue cutting-edge work in an often-encountered (idea category, research area) pair is thus arguably more valuable than is the ability to pursue cutting-edge work in a rarely encountered (idea category, research area) pair. The implicit assumption in this approach is that, even though it is not yet known which (idea category, research area) pairs will be the most important sources of future progress in biomedicine, the past is the best predictor of the future.

Because the overall scientific frontier position for a nation (the edge factor) is calculated as a weighted sum over its position across all (idea category, research area) pairs, the resulting measure for the nation reflects its overall capability across all of biomedicine, as opposed to only the nation's capabilities in areas where it has concentrated most of its own activities. Accordingly, the edge factor is high only if the country has significant capabilities across different areas of biomedicine; expertise in a narrow subset of biomedicine is not enough.

However, in a secondary analysis we show that the results are robust to the case when the edge factor is calculated using as weights each country's own number of papers that link to a given (idea category, research area) pair. Hence, the results from this alternative specification reflect the novelty of the work actually pursued by the nation – emphasizing more the novelty of the nation's work in areas where it publishes a lot – rather than the nation's capabilities across all of biomedicine.

Cells with missing observations. Not all locations have publications linked to every (idea category, research area) pair. The results are mainly insensitive to how such missing cells are handled. One reason for this is that while there are 13,394 (idea category, research area) pairs with positive weights, the 3,000 largest (idea category, research area) pairs as measured by their weight account for the vast majority (82%) of the total weight with most of them large enough for also the small and mid-sized nations to be active in them. In our main specification, we handle the cells with missing observations by replacing the nation's edge factor for that cell with the nation's weighted average across the other cells (those (idea category, research area) pairs for which the location does have publications linked to it). The weights used in this calculation are the same weights as discussed above. In an alternative specification, we replace cells with missing edge scores with 0 (the worst possible edge score). In another alternative specification, we replace cells with missing edge scores with 100 (by definition the average novelty score for every (idea category, research area) pair). In both cases the results are similar to the results for the main specification (see section 3.2).

### 2.8 Comparison of the Approach with the Approach Used in Closest Prior Work

The approach employed here has two main differences with the closest prior work (Packalen and Bhattacharya, 2017). In addition to the shift in substantive focus – from ranking journals to ranking nations – the present analysis is conducted at the contribution-level, with contribution defined as a link from a paper to an (idea category, research area) pair. By contrast, in this prior work, the analysis was conducted at the paper-level. That is, here the novelty score for an entity is calculated first at the contribution-level separately for each (idea category, research area) pair and the overall novelty score for the entity is then calculated as a weighted sum across these each (idea category, research area) pairs. By contrast, in the prior work (Packalen and Bhattacharya, 2017) the novelty score for an entity was calculated at the paper-level either without controlling for either the idea category or the research area, or by only controlling for the research in a manner that essentially uses as weights the entity's own involvement in the research area (in this prior work the research area was determined based on the appearance of 6-digit MeSH terms; the entity of interest in this prior work was a journal, here it is a nation).

The advantage of the approach pursued in the present analysis is thus not only that the present approach controls for the idea category but also that the present approach uses as weights the (idea category, research area) pair's overall importance in biomedicine (as measured by the total number of contributions linked to it). This yields a better reflection of an entity's capabilities in biomedicine compared to the case when the weights represent the distribution of the entity's own involvement across different areas of biomedicine.

## 3. Results

### 3.1 Main Results

Figure 1 shows the edge factor for each nation based on papers published during 2015-2016. The edge factor is normalized so that the average edge factor across all contributions is 100. An edge factor of 110 for a nation indicates that on average the nation's contributions build on relatively new ideas 10% more often compared to the average contribution in the same research area. Markers drawn in red (blue) indicate edge factors that are well above (well below) average. Markers drawn in gray indicate edge factors that are approximately average.

The results show that the United States and South Korea have the most advanced positions on the scientific frontier: scientists working in these nations build on cutting-edge ideas more often than do scientists in other locations. The propensity for novel science is well above average also in Singapore and Taiwan. Countries that come after these four countries have approximately average propensity for novel science. Such countries include China, Canada, most western European countries (including the United Kingdom and Germany), Australia, and South Africa. Other countries (including Turkey, India, Brazil, and Iran) come further behind – scientists in these countries have clearly below average propensities for novel work.

Confidence intervals and results for alternative specifications (shown in Table S4 (Web Appendix) and discussed in detail below in subsection 3.2) indicate that in most cases these results are robust. The one exception is Saudi Arabia, for which results from alternative specifications suggest a below average tendency for novel work.

Countries examined here thus have quite different propensities for work with newer ideas in biomedicine. This indicates that location continues to exert considerable influence on what kind of science is pursued. Furthermore, even developed nations are not on an equal footing in the pursuit of novel scientific work: in some developed nations scientists take advantage of opportunities created by the arrival of new ideas much more often than do scientists in other nations.

Figure 2 shows the change in the edge factor for each nation from the 1990s to present. South Korea, Taiwan, and China have leapfrogged most developed nations. Whereas the United States is still among the leaders, the relative positions of Switzerland and the United Kingdom are less advanced now than they were in the 1990s. Overall some convergence appears to have taken place as the lagging nations are no longer as far behind the leaders. This suggests that the world of ideas may have become somewhat flatter. Analysis of the edge factor by 5-year time periods (shown in Table S5 (Web Appendix)) indicates that most changes that occur are persistent. The changes thus reflect systematic changes in capabilities rather than merely year-to-year random variations.

In our approach, we compare each contribution only to other contributions that use ideas from the same idea category and are linked to the same research area (the 127 idea categories include “Amino Acid, Peptide, or Protein” and “Pharmacologic Substance”; the 125 research areas include “Biochemistry” and “Neoplasms”; see Table S1 and Table S2 for the full lists). Table 1 shows the edge factor separately for four groupings of idea categories: “Clinical and Anatomy”, “Drugs and Chemicals”, “Basic Science and Research Tools”, and “Miscellaneous”, and for three groupings of research areas: “Applied”, “Basic Science”, and “Other (Both Applied and Basic Science)”.

For most nations, the edge factor is similar across these groupings, suggesting that the pursuit of novel work is generally dependent on capabilities that some countries possess but others lack. One important exception is China. China’s contributions linked to the idea category grouping “Basic Science and Research Tools” now have the second highest propensity for novel work (after Singapore), but its contributions linked to idea category groupings “Clinical and Anatomy” and “Drugs and Chemicals” are well below average in terms of their novelty. This result serves to highlight an important feature of our approach: it can be used to reveal not just whether a nation is facing barriers in new idea adoption but where in the idea space those barriers lie.

While in Table 1 we divided papers to just three groups based on their research area (i.e. basic, applied, and other), it is important to note that the restriction to just three bins on this dimension was made for expositional purposes. The edge factors can also be reported separately for each of the 125 research areas (listed in Table S2). For example, funding agencies could use such field-level analyses of novelty across to determine in which scientific fields their own country is closest to the frontier and use this information (in conjunction with other relevant metrics) when deciding where to allocate limited research dollars.

### 3.2 Confidence Intervals and Results from Alternative Specifications

Table S4 (Web Appendix) shows the confidence intervals and results from a variety of alternative specifications. For ease of comparison, the results from the main specification are reported again in column (1d).

Confidence intervals reported column (1e) are constructed using a bootstrap method. We first generate each of 1000 artificial samples by re-sampling with replacement from the (idea category, research area) pairs until the total weighted number of observations (i.e. contributions) in each constructed sample is at least as large as the total weighted number of observations is in the original sample. Next, we calculate the edge factor for each nation in each constructed artificial sample. We then eliminate the largest 2.5% and the smallest 2.5% of the values in the edge factor distribution for each nation among these constructed bootstrapped samples. The extremes of remaining edge factor values form the 95% confidence interval for the edge factor of each nation.

The calculated confidence intervals indicate that scientists in the four top nations are clearly above average in their propensity to use new ideas, that scientists in most developed nations have approximately an average propensity to use new ideas, and that scientists in developing nations have a below average propensity to use new ideas.

The analysis reported in column (2) differs from the main specification in terms of how those (idea category, research area) pairs are treated for which a nation has no contributions linked to it: now the edge factor for such (idea category, research area) pairs are replaced with 0, reflecting the most pessimistic scenario about the nation's capabilities for that (idea category, research area) pair. By contrast, in the main specification these missing observations are replaced with the average edge score for the nation for (idea category, research area) pairs for which the nation does have observations. Comparison of the main results (column 1c) against the results in column (2) shows that while the edge factor decreases somewhat for the smaller nations (as expected), the results remain qualitatively unchanged. The fact that regardless of the approach the nations near the top of the rankings included also smaller and medium sized nations (in terms of their scientific output, as indicated by Table S3), and the fact that some larger countries are far down in the rankings (most notably India), demonstrate that the results are not driven solely by the size of a country's scientific workforce.

The analysis reported in column (3) differs from the main specification in how the weights for the edge factor for each (idea category, research area pair) are calculated. Here, for each country the weight for an (idea category, research area) pair is the country's own total number of research publications linked to the same (idea category, research area) pair. Thus, the overall edge factor is the same as the average of the nation's novelty scores across all of its contributions. By contrast, in the main specification weight for each (idea category, research area) pair is the same for all nations: it is the total number of papers linked to that (idea category, research area) pair. Comparison of the main results (column 1) against the results reported in column (3) shows that the results are robust to this alternative specification.

The analyses reported in columns (4-6) differ from the main specification in that the dummy variable indicating novelty of a contribution is now constructed using top 20%, top 10% and top 1% cutoffs. By contrast, in the main specification this dummy variable is constructed using the top 5% cutoff. Comparison of the main results (column 1) against results reported in columns (4-6) indicates that while the main results are qualitatively robust – leaders do better than laggards regardless of the measure – the relative position of the United States improves as one moves to a narrower cutoff (from 5% to 1%) and China's relative position improves when one moves to a wider cutoff (from 5% to 10% and 10% to 20%). A possible explanation is that countries may differ in terms of how many of their institutions are on the very edge of the frontier ("the bleeding edge"), so that some countries to fare better when novelty is calculated based on a narrower measure. For example, the U.S. may have many of the very top institutions in the world (in terms of their tendency to work with new ideas) but most of its institutions may be further down in the pack. In another country, such as China, institutions may be more homogenous in terms of the scientists' tendency to work with new ideas. The differences may also be driven by variation in where the new ideas are first born (the United States may be disproportionately the origin of new ideas – and thus receive a disproportionate share of the very first mentions of new terms – but scientists working in China may be relatively more eager to build on the new ideas).

The analysis reported in column (7) differs from the main specification in that now the cohort of each UMLS term is the year of the earliest mention of that term or any of its synonyms, with synonyms specified by the UMLS. In contrast, in the main specification the cohort year is the year of the earliest mention of the term itself. Comparison of the main results (column 1) against the results reported in column (7) shows that the conclusions from the main specification are robust in this way as well.

The analysis reported in column (8) differs from the main specification in that the analysis now includes all publications in MEDLINE as opposed to only regular research articles. The analysis reported in column (9) in turn differs from the main specification in that the analysis now includes also publications for which the text information on the title and abstract is less than 200 characters or more than 5000 characters – in the main specification such publications were excluded from the analysis. Comparison of the main results (column 1) against the results reported in columns (8) and (9) show that the results are robust also to these alternative specifications.

## 4. Discussion

While our results show that differences persist even among developed nations in their propensity to work with new ideas, the results do not reveal the specific mechanisms driving these differences. One potential driver of these cross-locational differences stems from the difficulty of working with new ideas. Because novel science is harder than conventional science, novel science is more dependent on interactions with colleagues. The fertility of these scientist interactions depends on factors such as the extent of complementary tacit knowledge that is embedded in people and is transferred to others in meetings (Lucas 2004; Lucas and Moll, 2014). Cross-national variation in the extent and depth of human capital investments can thus lead to cross-national variation in the tendency to adopt new ideas.

Of course, not all fruitful interactions are limited by location, as is evidenced by the fact that a quarter of science now involves international collaborations (Freeman, 2013; National Science Board, 2016). However, the rise of long-distance collaborations can also be a source of cross-national differences in new idea adoption: a nation can gain an advantage if its scientists can form distant collaborations relatively easily. In this regard, China's special relationship with the United States in science (Freeman and Huang, 2015) has likely helped propel it to the scientific frontier. An important topic for future work is to explore to what extent Chinese scientists working at the frontier started their work in the U.S. This link would potentially have major implications for other nations that are seeking to advance their position relative to the scientific frontier. A related topic worthy of future exploration is quantifying to what extent national borders still limit collaboration opportunities and what are the implications of collaboration barriers for each country in terms of the novelty of its scientific output.

Willingness to try out new ideas can vary by location also due to differences in scientist demographics. For example, given that early-career scientists are the most likely to work with new ideas (Packalen and Bhattacharya, 2016), and given that the increase in the extent of science in China is so recent and thus many of its scientists are early on their careers, the novelty of science in China may be driven in part by the youth of its scientists. Cross-national differences in new idea adoption and China's remarkably ability to leapfrog in this regard may also be driven in part by differences in incentives to pursue novel work: it has long been understood that nations without vested interests in existing technologies have an elevated incentive to explore new ideas (Brezis et al., 1993; Mokyr, 1994). Some of the variation in new idea adoption can also be driven by variation in where the ideas are first born, and by remaining delays in the spread of awareness about which new ideas exist.

Our results are consistent with findings from Hidalgo and Hausman (2009) which measured the complexity of each country's production structure based on its exports and found large differences in the capabilities of nations. Their analysis was motivated by the idea that each nation's capabilities determine the input varieties that it can fruitfully use in production. Our work, by contrast, is motivated by the idea that capabilities determine whether a nation's scientists can take advantage of the opportunities created by the arrival of new ideas. Moreover, whereas in this related work the complexity of goods production is measured indirectly based on exports, the edge factor is calculated directly based on the measured idea inputs. Common to these analyses is the belief that the capabilities of a nation affect which inputs it uses and both analyses are aimed at constructing new measures that reflect those capabilities.

Our finding that nations continue to differ in their ability to pursue novel science is in line with cross-country comparisons of scientific impact as measured by citations (Freeman, 2013; National Science Board, 2016). The ability to take advantage of scientific opportunities continues to vary across locations in spite of the "death of distance" phenomenon, because locational differences in capabilities persist as shown by Jones et al. (2008), Agrawal and Goldfarb (2008), Ding et al. (2010) and Packalen and Bhattacharya (2015).

But some aspects of our results also differ from the results obtained through traditional analyses of scientific productivity. Data on the tendency to produce highly cited papers point to the United States as a leader that remains far ahead of most western European nations and even further ahead of South Korea, Taiwan and China (Freeman, 2013; National Science Board, 2016; Freeman and Huang, 2015; Bornmann et al., 2017; Yu et al., 2014). Our analysis on the use of new ideas, by contrast, suggests that South Korea, Taiwan and China have caught up with western Europe and are now close to the United States in terms of their tendency to work with cutting-edge ideas. Moreover, we find that China is now a leader in favoring newer ideas when working with new basic science ideas and research tools.

The finding that some countries are among the leaders in terms of their edge factor but lag in terms of their impact is not surprising. Prior studies have found novelty and impact to correlate only imperfectly (Packalen and Bhattacharya, 2017; Wang et al., 2016; Lee et al., 2015). Moreover, work on an idea early – when the idea is still raw – may well have less impact than work that builds on more established ideas which properties are better understood. The early work on the idea is still crucial: it helps the idea develop and thus makes more significant advances possible. Moreover, countries investing heavily in novel science can reap significant benefits also for themselves from their focus: early work on an idea can help the country develop capabilities that enable it to take advantage of the later, more fertile, opportunities linked to the same idea.

## 5. Conclusion

Our analysis has shown that countries continue to differ in their ability to take advantage of cutting-edge ideas. Even across developed nations, sizable differences persist in terms of the tendency at which each nation's scientists build on recent advances. Hence, in spite of the arrival of modern communication technologies – which now facilitate almost instantaneous access to each new idea from almost anywhere in the world – the world of ideas is not yet flat. A likely explanation for this is that access to an idea does not guarantee that a scientist can take advantage of it in a productive way soon after its initial discovery. Instead, because new ideas are often raw and poorly understood, the ability to fruitfully build upon an idea depends on local factors such as daily interactions with colleagues, the training environment, and ready access to potential collaborators. Because these factors vary across locations but are beneficial or even necessary in helping scientists unlock the mysteries of new ideas, geographic differences in the tendency to exploit new ideas continue to persist.

Currently, the tendency to build on new ideas in biomedicine is highest in United States and South Korea. While China has leapfrogged most developed nations in terms of its overall tendency to work with new ideas, this progress has been uneven across idea types. In terms of applying basic science ideas, China has already caught up with the leaders, but in terms of applying new clinical knowledge, China remains below average. This result highlights an important benefit of the approach developed here: the approach can be used to show not only whether differences in new idea adoption persist, but also where in the idea space any remaining barriers lie.

An important direction for future work is to examine what are the best approaches for overcoming barriers in idea adoption. For example, it would be useful to uncover whether the scientific frontier is best reached through an approach where resources are first directed to a handful of fields or universities or through a more diversified approach, and whether smaller countries can elect to specialize in narrow areas or whether the unpredictability of where in the ideas space important future advances come from render it necessary for even smaller countries to develop and maintain broad capabilities so that they can take early advantage of new advances no matter where in the idea space those yet unforeseen new advances are born.

Extending the analysis to patent data would also be useful as it would facilitate exploration of links between scientific frontier positions and technological frontier positions. Such analysis could reveal to what extent funding scientists working near the edge of the scientific frontier is a pre-condition for a country to obtain the capability to pursue inventions near the edge of the technological frontier – inventions that build on recent scientific and technological advances.

In analyses of science, the metric introduced here – the edge factor – holds considerable potential beyond cross-national comparisons. Ever since Garfield (1955, 1972), the focus in empirical analyses of knowledge production has been on measuring influence. This focus on impact in science policy has recently been decried by many, including Alberts (2013) and Osterloh and Frey (2015). For the theory of knowledge production implies that science policy decisions should be guided by not just the influence of scientific work but also what kind of science is being pursued – novel or conventional (Kuhn 1962; Besancenot and Vranceanu 2015). Because new ideas are raw when they are first born, they need the attention and revision by many scientists to mature into useful advances. But such work is risky, and without explicit incentives for novel work, too many choose to pursue well-trodden research paths in areas where many other scientists also work and thus the prospects of receiving citations are better. A potential consequence of excessive focus on impact is thus that science becomes stagnant, a phenomenon which appears to have already occurred in biomedicine (Rzhetzky et al., 2015).

Of course, given the absence of metrics that capture novelty, the obsession with impact has been inevitable. The edge factor is a valuable tool for this very reason. By providing an approach that can be used to measure the novelty of scientific work of a nation, a journal, an institution, or even an individual scientist, the edge factor allows university administrators and funding agencies to re-structure their reward systems so that scientists are rewarded based on not just the impact of their work but also the novelty of their work. When scientists are rewarded based on *both* the impact and novelty of their work, more scientists can be expected to pursue novel research paths, leading to healthier, less stagnant science.

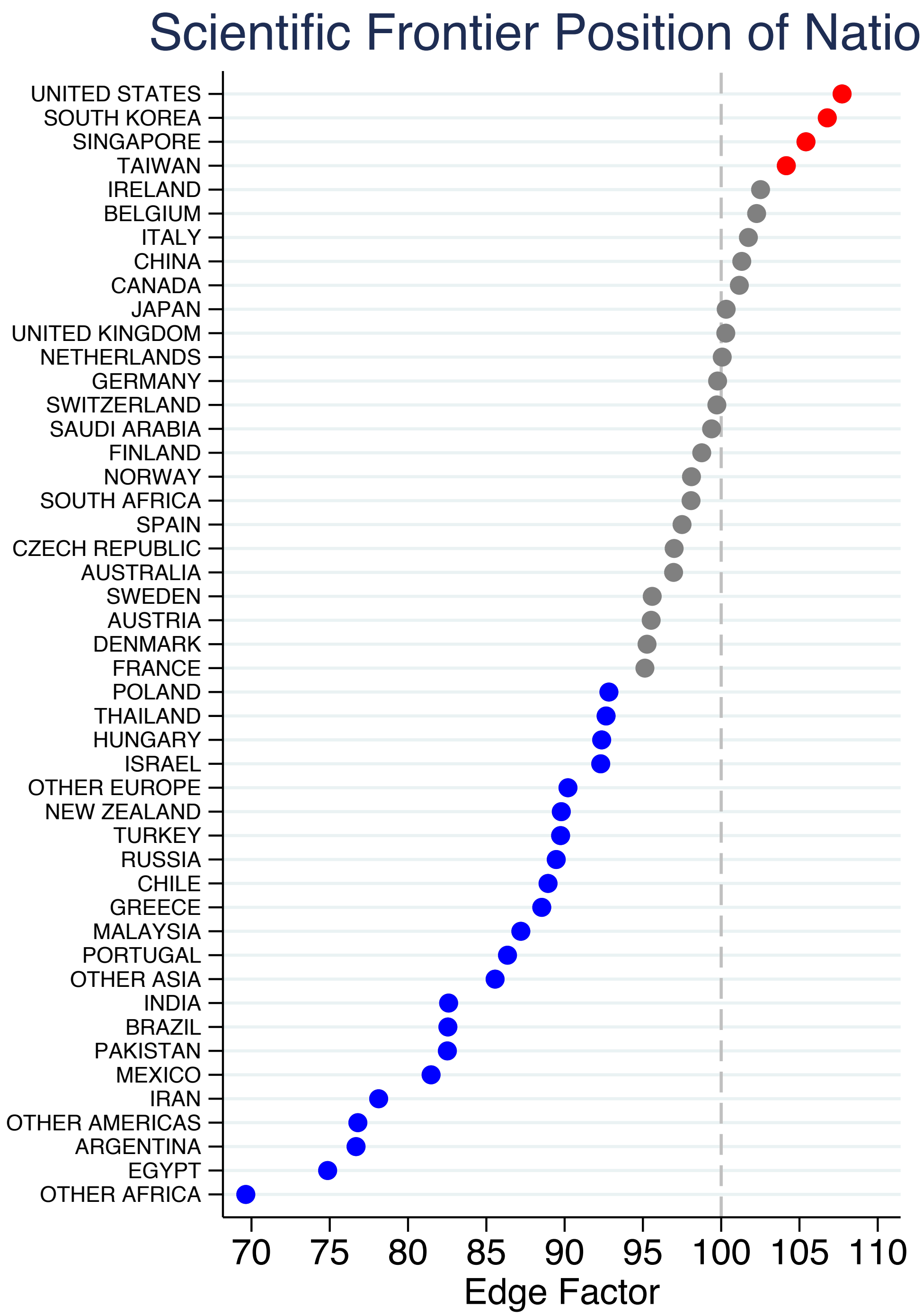

**Figure 1.** Overall Scientific Frontier Position by Location. Edge factors are calculated using text analysis and data on biomedical research papers published during 2015-2016. Scatter points are colored to indicate edge factors that are well above average (red), about average (grey), and well below average (blue). An edge factor above 100 indicates an above average tendency for work that builds on relatively new ideas (a contribution is considered novel if it is in the top 5% by the age of the newest idea it builds upon; the comparison group for each contribution is all other papers published in the same year and linked to the same (idea category, research area) pair).

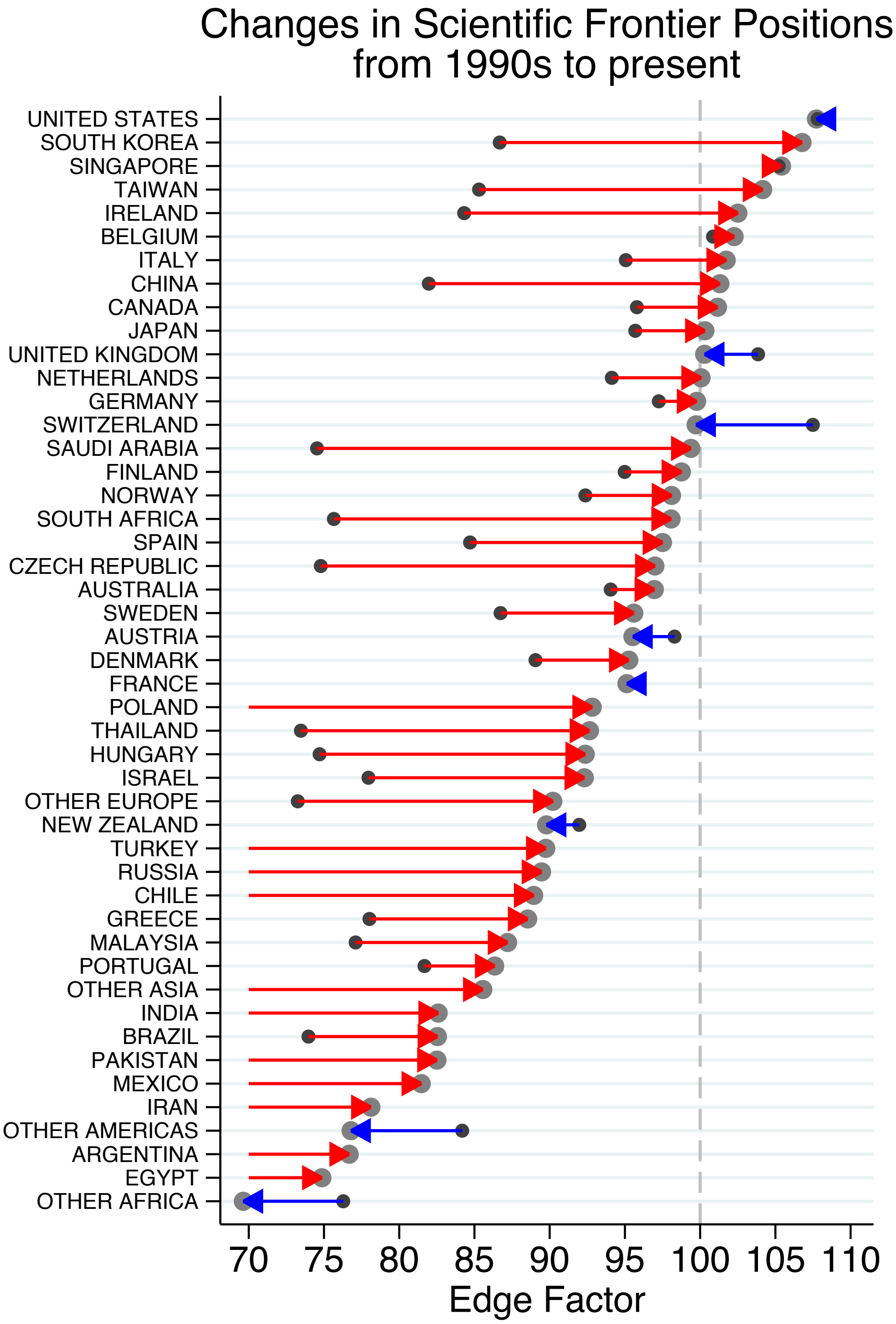

**Figure 2.** Change in Scientific Frontier Position by Location from 1990s to present. Edge factors are calculated using text analysis and data on biomedical research papers published during 1990-1999 and 2015-2016. Smaller scatter points indicate the edge factors for 1990s, larger points for 2015-6. Red (blue) arrows indicate edge factors that increased (decreased) from 1990s to 2015-2016. An edge factor above 100 indicates an above average tendency for work that builds on relatively new ideas (a contribution is considered novel if it is in the top 5% by the age of the newest idea it builds upon; the comparison group for each contribution is all other papers published in the same year and linked to the same (idea category, research area) pair).

**Table 1.** Edge Factors by Idea Category Type and by Research Area Type.

| (1a) | (1b) | (1c) | (2a) | (2b) | (2c) | (2d) | (3a) | (3b) | (3c) |
|---|---|---|---|---|---|---|---|---|---|
| Location | Number of Contributions | 2015-6 | Clinical and Anatomy | Drugs and Chemi-cals | Basic Science and Research Tools | Miscel-laneous | Applied | Basic Science | Other (Both Applied and Basic Science) |
| UNITED STATES | 2853661 | 108 | 105 | 121 | 110 | 105 | 106 | 108 | 108 |
| SOUTH KOREA | 374227 | 107 | 111 | 103 | 105 | 105 | 103 | 109 | 106 |
| SINGAPORE | 52541 | 105 | 109 | 106 | 115 | 108 | 108 | 110 | 112 |
| TAIWAN | 177229 | 104 | 99 | 100 | 105 | 105 | 100 | 101 | 107 |
| IRELAND | 39495 | 103 | 107 | 88 | 108 | 98 | 99 | 101 | 108 |
| BELGIUM | 95644 | 102 | 109 | 120 | 99 | 98 | 103 | 106 | 102 |
| ITALY | 384029 | 102 | 105 | 117 | 94 | 101 | 99 | 103 | 103 |
| CHINA | 1734035 | 101 | 95 | 88 | 113 | 101 | 102 | 100 | 103 |
| CANADA | 375846 | 101 | 99 | 94 | 102 | 105 | 101 | 99 | 105 |
| JAPAN | 554589 | 100 | 104 | 106 | 103 | 92 | 94 | 104 | 101 |
| UNITED KINGDOM | 494917 | 100 | 100 | 105 | 100 | 100 | 98 | 102 | 100 |
| NETHERLANDS | 233631 | 100 | 106 | 87 | 100 | 97 | 95 | 103 | 100 |
| GERMANY | 539888 | 100 | 95 | 112 | 104 | 97 | 96 | 102 | 100 |
| SWITZERLAND | 123779 | 100 | 97 | 118 | 105 | 93 | 94 | 102 | 102 |
| SAUDI ARABIA | 34855 | 99 | 96 | 84 | 91 | 96 | 90 | 92 | 98 |
| FINLAND | 59534 | 99 | 96 | 78 | 106 | 96 | 89 | 99 | 102 |
| NORWAY | 63699 | 98 | 99 | 88 | 103 | 98 | 97 | 97 | 104 |
| SOUTH AFRICA | 43179 | 98 | 107 | 81 | 76 | 98 | 100 | 92 | 89 |
| SPAIN | 278504 | 98 | 98 | 96 | 96 | 99 | 92 | 99 | 101 |
| CZECH REPUBLIC | 44024 | 97 | 97 | 86 | 95 | 92 | 95 | 97 | 89 |
| AUSTRALIA | 320955 | 97 | 99 | 94 | 95 | 97 | 96 | 95 | 100 |
| SWEDEN | 138949 | 96 | 94 | 102 | 97 | 93 | 91 | 98 | 96 |
| AUSTRIA | 65039 | 96 | 94 | 103 | 100 | 94 | 91 | 100 | 96 |
| DENMARK | 105066 | 95 | 98 | 91 | 96 | 96 | 92 | 95 | 101 |
| FRANCE | 305065 | 95 | 96 | 101 | 95 | 93 | 91 | 97 | 95 |
| POLAND | 113074 | 93 | 95 | 85 | 83 | 91 | 95 | 89 | 84 |
| THAILAND | 40080 | 93 | 95 | 79 | 73 | 94 | 91 | 81 | 92 |
| HUNGARY | 28574 | 92 | 84 | 94 | 93 | 85 | 88 | 89 | 87 |
| ISRAEL | 76781 | 92 | 96 | 78 | 95 | 90 | 88 | 95 | 94 |
| OTHER EUROPE | 107712 | 90 | 91 | 79 | 82 | 84 | 85 | 83 | 88 |
| NEW ZEALAND | 38946 | 90 | 90 | 111 | 88 | 93 | 92 | 95 | 90 |
| TURKEY | 157825 | 90 | 101 | 79 | 69 | 93 | 87 | 85 | 91 |
| RUSSIA | 51759 | 89 | 77 | 91 | 93 | 85 | 86 | 85 | 84 |
| CHILE | 23794 | 89 | 98 | 76 | 75 | 91 | 95 | 86 | 83 |
| GREECE | 46646 | 89 | 95 | 88 | 76 | 86 | 86 | 87 | 86 |
| MALAYSIA | 37997 | 87 | 87 | 64 | 81 | 90 | 87 | 83 | 83 |
| PORTUGAL | 65523 | 86 | 92 | 84 | 85 | 93 | 92 | 91 | 85 |
| OTHER ASIA | 60973 | 86 | 87 | 84 | 73 | 81 | 85 | 78 | 82 |
| INDIA | 291215 | 83 | 83 | 70 | 73 | 95 | 86 | 83 | 80 |
| BRAZIL | 274896 | 83 | 83 | 74 | 71 | 96 | 83 | 82 | 82 |
| PAKISTAN | 27511 | 83 | 78 | 93 | 74 | 80 | 81 | 79 | 77 |
| MEXICO | 54997 | 81 | 81 | 72 | 72 | 86 | 82 | 76 | 80 |
| IRAN | 121035 | 78 | 87 | 62 | 68 | 82 | 77 | 76 | 81 |
| OTHER AMERICAS | 30787 | 77 | 82 | 76 | 64 | 96 | 87 | 78 | 78 |
| ARGENTINA | 40775 | 77 | 85 | 76 | 69 | 86 | 82 | 79 | 78 |
| EGYPT | 48649 | 75 | 85 | 75 | 58 | 81 | 76 | 75 | 74 |
| OTHER AFRICA | 90041 | 70 | 87 | 57 | 55 | 73 | 76 | 68 | 72 |

Notes to Table 1:

All numbers in columns 1b and 1c are calculated based on papers published during 2015-2016. Numbers in columns 2a-d and columns 3a-3c are also calculated based on papers published during 2015-2016 for all countries for which the number of contributions reported in column 1c is at least 200000. For countries that fall below this threshold, the numbers in columns 2a-d and columns 3a-3c are calculated based on papers published during 2010-2016 (in order to decrease the variability of the edge factors reported in these columns).

Column 1a: Location.

Column 1b: Number of contributions based on which the edge factor in column (1c) is calculated. A contribution is defined as a link from a paper to an (idea category, research area) pair. A paper can link to multiple (idea category, research area) pairs because a paper can mention UMLS terms from multiple idea categories, and because a UMLS term can be linked to multiple idea categories, and because a paper may be linked to multiple research areas.

Column 1c: Edge factor for the baseline specification.

Columns 2a-d: Edge factors for each of the four idea category groups "Clinical and Anatomy", "Drugs and Chemicals", "Basic Science and Research Tools", and "Miscellaneous". See Table S1 for which idea categories (as represented by UMLS categories for UMLS terms) are included in each idea category group.

Columns 3a-c: Edge factors for each of the three types of research areas: "Applied", "Basic Science", and Other (Both Applied and Basic Science)". See Table S1 for which research areas (represented by journal categories) are included in each of these three research area types.

Web Appendix:

# Edge Factors: Scientific Frontier Positions of Nations

Mikko Packalen

June 18, 2018

**This PDF file includes:**

Tables S1-S5
Figures S1-S5

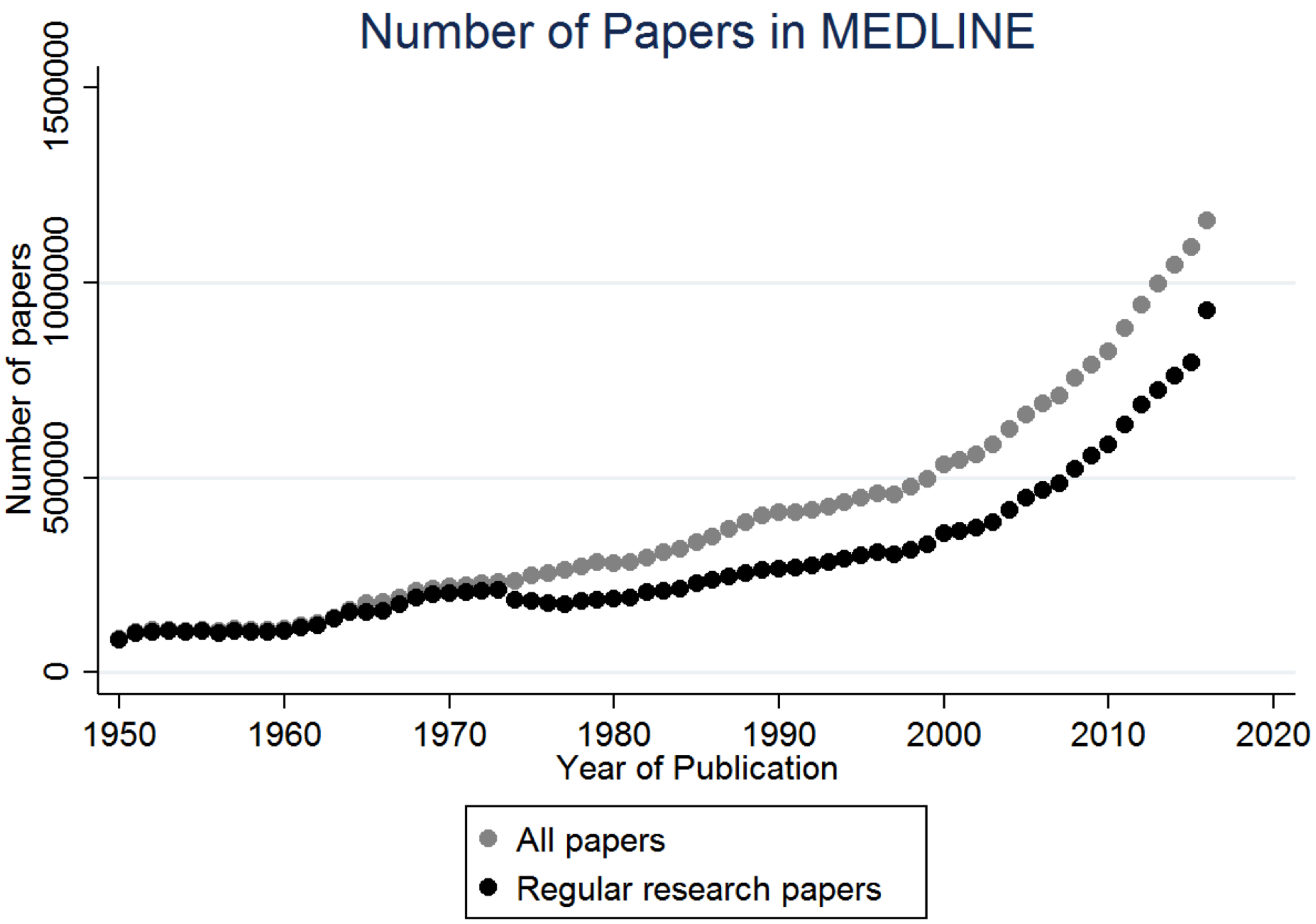

**Figure S1.** Number of papers per year in the MEDLINE database. Even after limiting the analysis to regular research papers (thereby excluding news items, editorials, etc.), the database includes mullions of papers for even the earlier decades allowing us to obtain an informative estimate of when each idea was new to the biomedical literature.

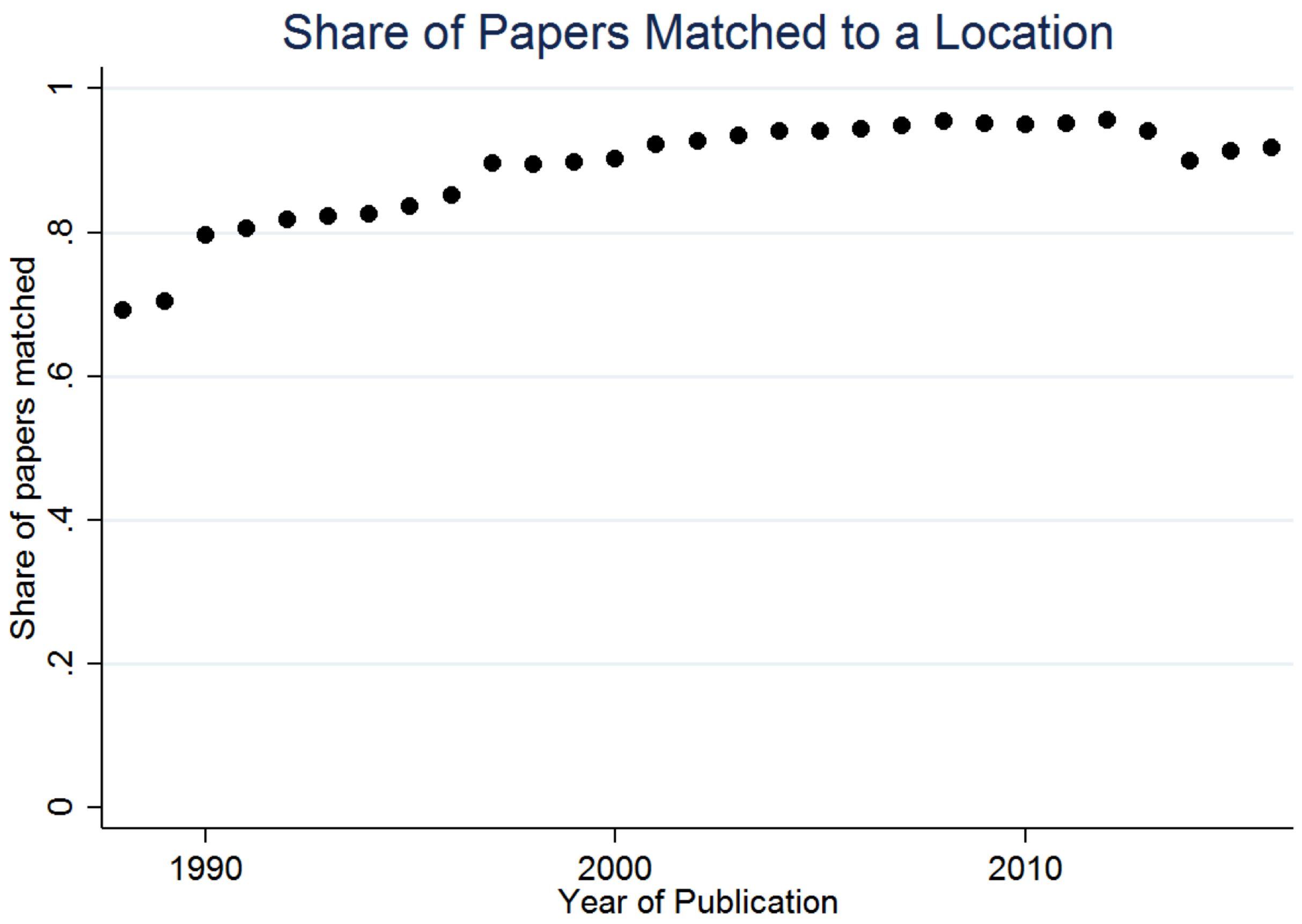

**Figure S2.** Share of papers matched to a location. The match rate is reasonably high even for 1990s. The decrease in the rate of matched papers in recent years is due to the fact that for those years some of the affiliation strings in MEDLINE include the affiliation string for multiple authors. The form of such entries makes it more difficult to match those papers to a country.

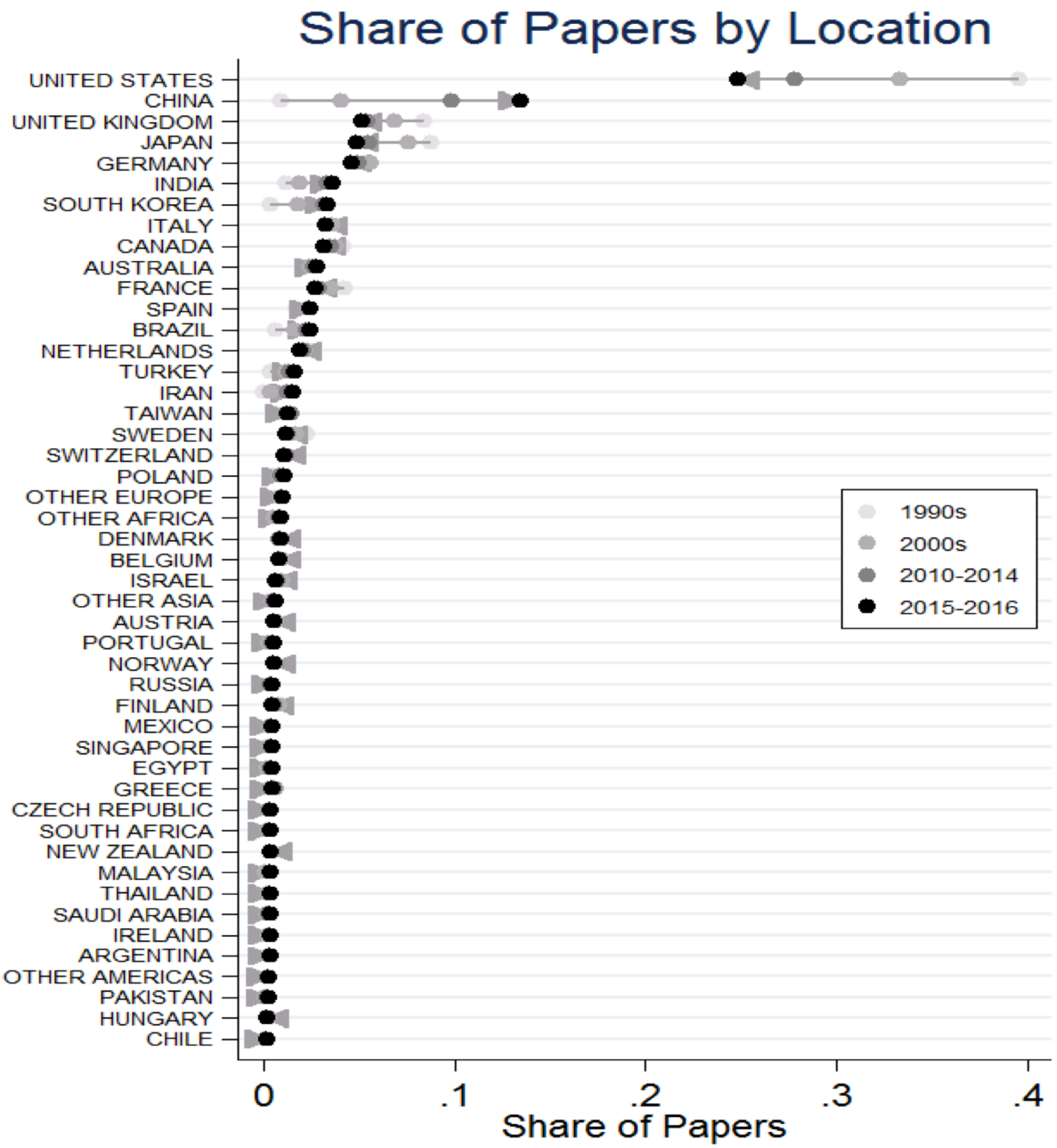

**Figure S3.** Share of papers by location. The U.S., U.K., Japan, Germany, and Italy have been among near the top in terms of extent of biomedical research production throughout the sample period, while China, India, and South Korea have become big centers of biomedical research production during the last two decades.

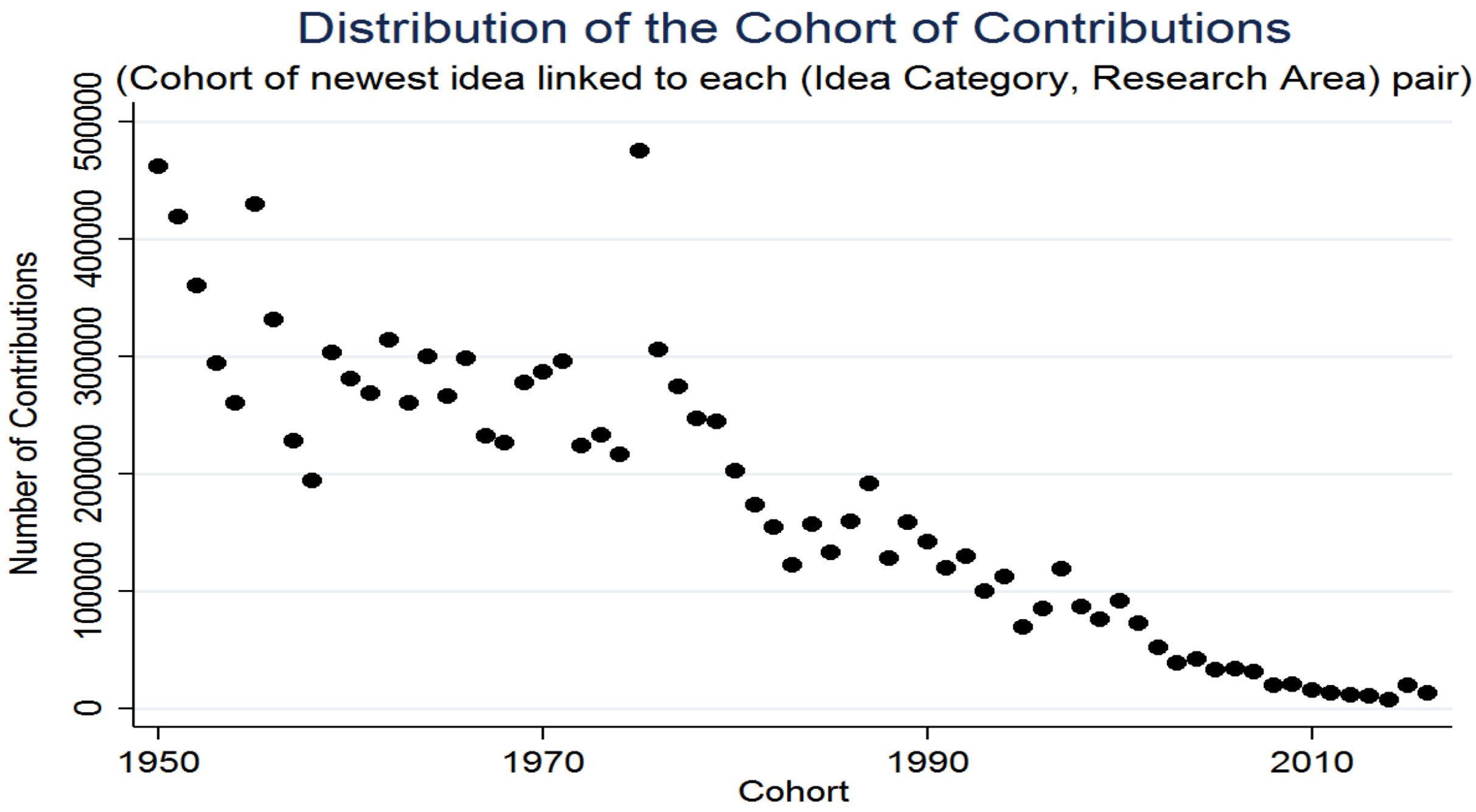

**Figure S4.** Distribution of the Cohort of Contributions. The number of contributions with cohort "1975" is disproportionately high because the comprehensive coverage of article abstracts in MEDLINE begins in 1975 (and thus a disproportionate number of terms are assigned cohort 1975 by our approach).

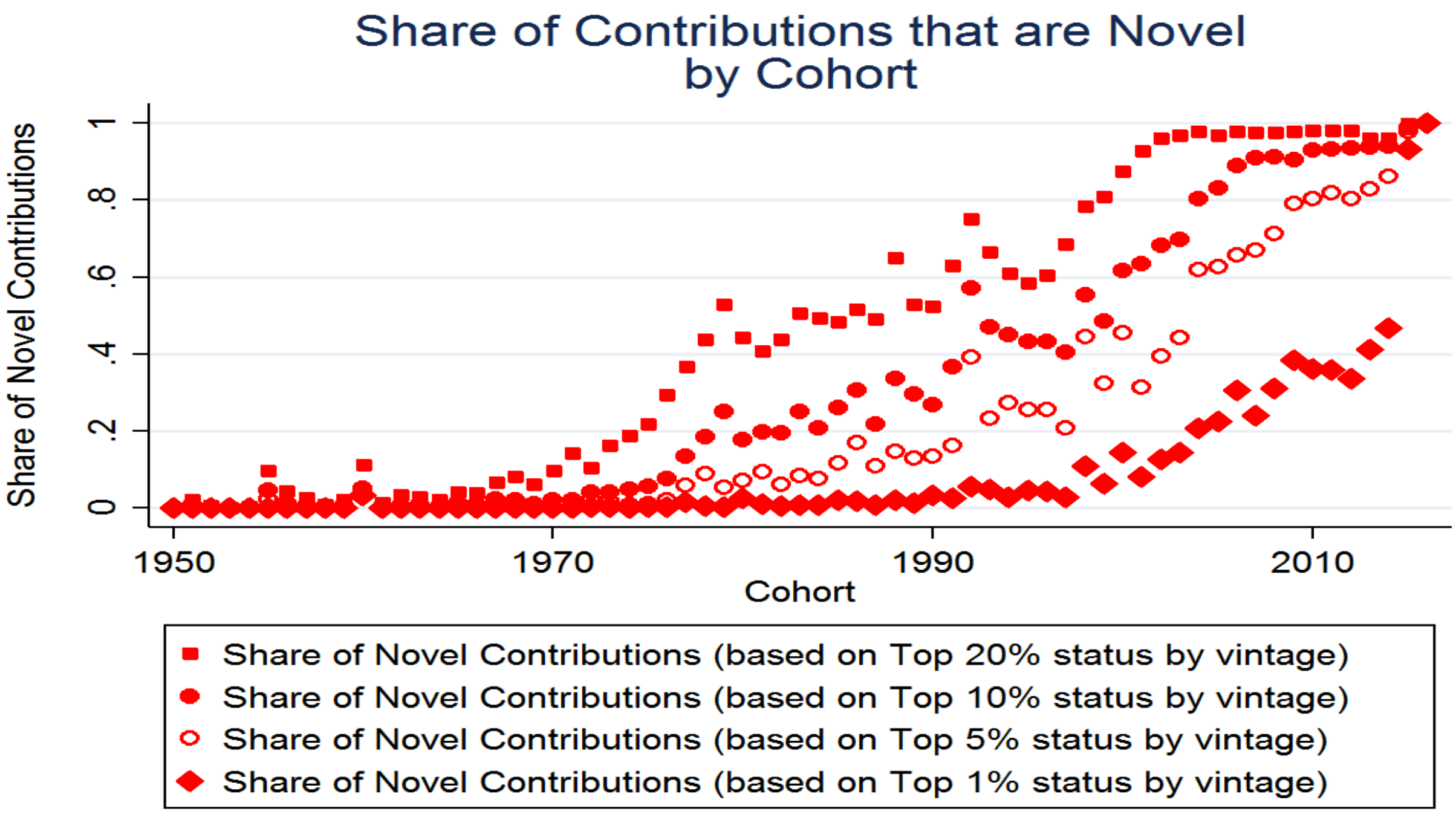

**Figure S5.** Share of Novel Contributions by Cohort. When the top 5% cutoff is used, then for all post-2004 cohorts the majority of contributions with that cohort are deemed novel by our approach, and for all pre-2004 cohorts at most a minority of contributions are deemed novel by our approach. Contributions with a very early cohort are never novel and contributions with the latest possible cohort ("2016") are all novel. Contributions with a cohort between these extremes are sometimes novel and other times not. This is because novelty is calculated by comparing the vintage of a contribution to the vintage of other contributions linked to the same (idea category, research area) pair. Hence, the cutoff cohort for novel contributions varies across (idea category, research area) pairs.

**Table S1. Number of Links from Papers to Each Research Area.**

| Research Area (Journal Category) | Links | Research Area Group |
|---|---|---|
| Medicine | 669881 | Other (Both Applied and Basic Science) |
| Science | 599469 | Basic Science |
| Neoplasms | 524947 | Other (Both Applied and Basic Science) |
| Biochemistry | 499331 | Basic Science |
| Molecular Biology | 470936 | Basic Science |
| Neurology | 404341 | Other (Both Applied and Basic Science) |
| Chemistry | 381274 | Basic Science |
| Pharmacology | 283555 | Other (Both Applied and Basic Science) |
| Biology | 264824 | Basic Science |
| Cell Biology | 256341 | Basic Science |
| General Surgery | 247670 | Applied |
| Environmental Health | 233734 | Other (Both Applied and Basic Science) |
| Allergy and Immunology | 216658 | Basic Science |
| Microbiology | 196761 | Basic Science |
| Cardiology | 192365 | Applied |
| Biomedical Engineering | 191269 | Other (Both Applied and Basic Science) |
| Biotechnology | 184539 | Basic Science |
| Biophysics | 174613 | Basic Science |
| Vascular Diseases | 174607 | Other (Both Applied and Basic Science) |
| Physiology | 172288 | Other (Both Applied and Basic Science) |
| Public Health | 163958 | Applied |
| Pediatrics | 163047 | Applied |
| Nutritional Sciences | 161807 | Other (Both Applied and Basic Science) |
| Toxicology | 158233 | Other (Both Applied and Basic Science) |
| Psychiatry | 155919 | Applied |
| Gastroenterology | 129905 | Other (Both Applied and Basic Science) |
| Endocrinology | 127616 | Other (Both Applied and Basic Science) |
| Psychology | 122978 | Applied |
| Genetics | 121180 | Basic Science |
| Nursing | 116035 | Applied |
| Ophthalmology | 112186 | Other (Both Applied and Basic Science) |
| Chemistry Techniques, Analytical | 107968 | Other (Both Applied and Basic Science) |
| Pulmonary Medicine | 103534 | Applied |
| Orthopedics | 102817 | Applied |
| Diagnostic Imaging | 101475 | Applied |
| Dentistry | 101252 | Applied |
| Pathology | 99477 | Other (Both Applied and Basic Science) |
| Metabolism | 98272 | Other (Both Applied and Basic Science) |
| Communicable Diseases | 96646 | Other (Both Applied and Basic Science) |
| Therapeutics | 95906 | Other (Both Applied and Basic Science) |
| Veterinary Medicine | 95394 | Other (Both Applied and Basic Science) |
| Radiology | 95332 | Applied |
| Behavioral Sciences | 91608 | Applied |
| Brain | 90578 | Other (Both Applied and Basic Science) |
| Nanotechnology | 87781 | Basic Science |
| Hematology | 87705 | Other (Both Applied and Basic Science) |
| Geriatrics | 82363 | Applied |
| Botany | 81144 | Other (Both Applied and Basic Science) |
| Physics | 80223 | Other (Both Applied and Basic Science) |
| Gynecology | 76024 | Applied |
| Genetics, Medical | 75252 | Other (Both Applied and Basic Science) |
| Health Services | 74519 | Applied |
| Psychophysiology | 70549 | Applied |
| Obstetrics | 69653 | Applied |
| Virology | 68388 | Basic Science |
| Technology | 66620 | Other (Both Applied and Basic Science) |
| Urology | 65771 | Applied |
| Reproductive Medicine | 63949 | Other (Both Applied and Basic Science) |
| Drug Therapy | 63105 | Other (Both Applied and Basic Science) |
| Zoology | 62592 | Other (Both Applied and Basic Science) |
| Medical Informatics | 61238 | Applied |
| Transplantation | 59013 | Other (Both Applied and Basic Science) |
| Health Services Research | 53889 | Applied |
| Traumatology | 53820 | Applied |
| Nephrology | 51255 | Other (Both Applied and Basic Science) |
| Physical and Rehabilitation Medicine | 50756 | Applied |
| Rheumatology | 50730 | Other (Both Applied and Basic Science) |
| Dermatology | 49397 | Other (Both Applied and Basic Science) |
| Epidemiology | 48742 | Applied |
| Social Sciences | 47935 | Applied |
| Sports Medicine | 45323 | Applied |
| Tropical Medicine | 44027 | Other (Both Applied and Basic Science) |
| Otolaryngology | 43871 | Applied |
| Computational Biology | 43373 | Basic Science |
| Radiotherapy | 42859 | Applied |
| Parasitology | 40296 | Other (Both Applied and Basic Science) |
| Substance-Related Disorders | 39722 | Applied |
| Complementary Therapies | 39646 | Other (Both Applied and Basic Science) |
| Anti-Infective Agents | 38006 | Basic Science |
| Neurosurgery | 37052 | Applied |
| Acquired Immunodeficiency Syndrome | 34489 | Other (Both Applied and Basic Science) |
| Nuclear Medicine | 33864 | Other (Both Applied and Basic Science) |
| Education | 33318 | Applied |
| Emergency Medicine | 32612 | Applied |
| Critical Care | 32054 | Applied |
| Anesthesiology | 31729 | Applied |
| Perinatology | 31395 | Applied |
| Clinical Laboratory Techniques | 30107 | Other (Both Applied and Basic Science) |
| Embryology | 29322 | Other (Both Applied and Basic Science) |
| Pharmacy | 28546 | Other (Both Applied and Basic Science) |
| Palliative Care | 27801 | Applied |
| Psychopharmacology | 27246 | Applied |
| Internal Medicine | 24728 | Applied |
| Occupational Medicine | 20677 | Applied |
| Statistics as Topic | 20049 | Applied |
| Antineoplastic Agents | 19027 | Other (Both Applied and Basic Science) |
| Primary Health Care | 18607 | Applied |
| Jurisprudence | 17346 | Other (Both Applied and Basic Science) |
| Histology | 15443 | Basic Science |
| Hospitals | 14323 | Other (Both Applied and Basic Science) |
| Audiology | 13697 | Applied |
| Sexually Transmitted Diseases | 11994 | Other (Both Applied and Basic Science) |
| Anatomy | 11926 | Other (Both Applied and Basic Science) |
| Ethics | 11039 | Applied |
| Bacteriology | 10395 | Basic Science |
| Speech-Language Pathology | 10150 | Applied |
| Women's Health | 9976 | Applied |
| Histocytochemistry | 8834 | Basic Science |
| Chemistry, Clinical | 8529 | Other (Both Applied and Basic Science) |
| Forensic Sciences | 8528 | Other (Both Applied and Basic Science) |
| Military Medicine | 7359 | Applied |
| Orthodontics | 5106 | Applied |
| Anthropology | 4340 | Applied |
| Laboratory Animal Science | 3784 | Other (Both Applied and Basic Science) |
| Vital Statistics | 3672 | Other (Both Applied and Basic Science) |
| History of Medicine | 3580 | Applied |
| Disaster Medicine | 3491 | Applied |
| Teratology | 1938 | Other (Both Applied and Basic Science) |
| Podiatry | 1671 | Applied |
| Aerospace Medicine | 1591 | Applied |
| Family Planning Services | 1396 | Applied |
| Chiropractic | 648 | Applied |
| Osteopathic Medicine | 385 | Applied |
| Library Science | 226 | Applied |

**Table S2. Number of Links from Papers to Each Idea Category.**

| Idea Category | Links | Idea Category Group |
|---|---|---|
| Finding | 606119 | Clinical and Anatomy |
| Amino Acid, Peptide, or Protein | 529613 | Basic Science and Research Tools |
| Pharmacologic Substance | 527933 | Drugs and Chemicals |
| Quantitative Concept | 495874 | Miscellaneous |
| Intellectual Product | 485671 | Miscellaneous |
| Laboratory Procedure | 478481 | Clinical and Anatomy |
| Gene or Genome | 470078 | Basic Science and Research Tools |
| Research Activity | 380742 | Basic Science and Research Tools |
| Therapeutic or Preventive Procedure | 374185 | Clinical and Anatomy |
| Disease or Syndrome | 353348 | Clinical and Anatomy |
| Molecular Function | 303528 | Basic Science and Research Tools |
| Functional Concept | 289601 | Miscellaneous |
| Clinical Attribute | 282872 | Clinical and Anatomy |
| Diagnostic Procedure | 261596 | Clinical and Anatomy |
| Manufactured Object | 244287 | Miscellaneous |
| Qualitative Concept | 239603 | Miscellaneous |
| Cell Function | 231778 | Basic Science and Research Tools |
| Genetic Function | 202810 | Basic Science and Research Tools |
| Organic Chemical | 200841 | Drugs and Chemicals |
| Mental Process | 184553 | Clinical and Anatomy |
| Health Care Activity | 182871 | Clinical and Anatomy |
| Cell | 172497 | Basic Science and Research Tools |
| Idea or Concept | 166488 | Miscellaneous |
| Nucleic Acid, Nucleoside, or Nucleotide | 155993 | Basic Science and Research Tools |
| Spatial Concept | 140830 | Miscellaneous |
| Molecular Biology Research Technique | 135534 | Basic Science and Research Tools |
| Neoplastic Process | 125951 | Clinical and Anatomy |
| Body Part, Organ, or Organ Component | 121709 | Clinical and Anatomy |
| Temporal Concept | 121231 | Miscellaneous |
| Medical Device | 119522 | Clinical and Anatomy |
| Biomedical Occupation or Discipline | 117883 | Miscellaneous |
| Cell Component | 113613 | Basic Science and Research Tools |
| Population Group | 112704 | Miscellaneous |
| Pathologic Function | 108290 | Clinical and Anatomy |
| Professional or Occupational Group | 106552 | Miscellaneous |
| Activity | 97839 | Miscellaneous |
| Mental or Behavioral Dysfunction | 88458 | Clinical and Anatomy |
| Indicator, Reagent, or Diagnostic Aid | 84232 | Clinical and Anatomy |
| Organ or Tissue Function | 80670 | Clinical and Anatomy |
| Plant | 79846 | Miscellaneous |
| Natural Phenomenon or Process | 78537 | Basic Science and Research Tools |
| Educational Activity | 76874 | Clinical and Anatomy |
| Biologically Active Substance | 70188 | Drugs and Chemicals |
| Sign or Symptom | 69915 | Clinical and Anatomy |
| Eukaryote | 69229 | Basic Science and Research Tools |
| Bacterium | 68427 | Basic Science and Research Tools |
| Cell or Molecular Dysfunction | 65978 | Basic Science and Research Tools |
| Hazardous or Poisonous Substance | 62697 | Drugs and Chemicals |
| Laboratory or Test Result | 61984 | Clinical and Anatomy |
| Injury or Poisoning | 61553 | Clinical and Anatomy |
| Conceptual Entity | 61528 | Miscellaneous |
| Social Behavior | 60906 | Miscellaneous |
| Mammal | 57221 | Miscellaneous |
| Organism Function | 55354 | Basic Science and Research Tools |
| Biomedical or Dental Material | 55315 | Drugs and Chemicals |
| Organism Attribute | 54953 | Miscellaneous |
| Virus | 51368 | Basic Science and Research Tools |
| Occupation or Discipline | 50613 | Miscellaneous |
| Individual Behavior | 48054 | Miscellaneous |
| Body Location or Region | 47276 | Clinical and Anatomy |
| Health Care Related Organization | 46763 | Miscellaneous |
| Classification | 46324 | Miscellaneous |
| Nucleotide Sequence | 45571 | Basic Science and Research Tools |
| Occupational Activity | 45153 | Miscellaneous |
| Phenomenon or Process | 44693 | Basic Science and Research Tools |
| Element, Ion, or Isotope | 43951 | Basic Science and Research Tools |
| Physiologic Function | 43616 | Clinical and Anatomy |
| Geographic Area | 40627 | Miscellaneous |
| Experimental Model of Disease | 38736 | Clinical and Anatomy |
| Amino Acid Sequence | 38461 | Basic Science and Research Tools |
| Machine Activity | 36253 | Miscellaneous |
| Tissue | 36151 | Basic Science and Research Tools |
| Immunologic Factor | 35682 | Basic Science and Research Tools |
| Organism | 32501 | Basic Science and Research Tools |
| Inorganic Chemical | 27299 | Drugs and Chemicals |
| Animal | 25586 | Miscellaneous |
| Food | 24315 | Miscellaneous |
| Age Group | 23986 | Miscellaneous |
| Daily or Recreational Activity | 23540 | Miscellaneous |
| Chemical Viewed Functionally | 21956 | Drugs and Chemicals |
| Fish | 21900 | Miscellaneous |
| Family Group | 21721 | Miscellaneous |
| Biologic Function | 21506 | Basic Science and Research Tools |
| Substance | 20164 | Basic Science and Research Tools |
| Group | 19705 | Miscellaneous |
| Body Space or Junction | 18744 | Clinical and Anatomy |
| Congenital Abnormality | 18684 | Clinical and Anatomy |
| Clinical Drug | 17971 | Drugs and Chemicals |
| Fungus | 17966 | Basic Science and Research Tools |
| Research Device | 17100 | Basic Science and Research Tools |
| Governmental or Regulatory Activity | 16215 | Miscellaneous |
| Body Substance | 16146 | Basic Science and Research Tools |
| Chemical Viewed Structurally | 15715 | Drugs and Chemicals |
| Chemical | 14247 | Drugs and Chemicals |
| Patient or Disabled Group | 13280 | Miscellaneous |
| Organization | 11961 | Miscellaneous |
| Receptor | 11335 | Basic Science and Research Tools |
| Human-caused Phenomenon or Process | 10689 | Basic Science and Research Tools |
| Bird | 9043 | Miscellaneous |
| Acquired Abnormality | 9040 | Clinical and Anatomy |
| Regulation or Law | 8969 | Miscellaneous |
| Anatomical Abnormality | 8896 | Clinical and Anatomy |
| Environmental Effect of Humans | 8268 | Miscellaneous |
| Body System | 7871 | Clinical and Anatomy |
| Group Attribute | 7303 | Miscellaneous |
| Behavior | 6400 | Miscellaneous |
| Embryonic Structure | 5498 | Basic Science and Research Tools |
| Professional Society | 4272 | Miscellaneous |
| Event | 3825 | Miscellaneous |
| Reptile | 3289 | Miscellaneous |
| Hormone | 2983 | Drugs and Chemicals |
| Self-help or Relief Organization | 2927 | Miscellaneous |
| Archaeon | 2620 | Basic Science and Research Tools |
| Vitamin | 2514 | Drugs and Chemicals |
| Language | 2422 | Miscellaneous |
| Anatomical Structure | 2293 | Clinical and Anatomy |
| Physical Object | 2017 | Miscellaneous |
| Amphibian | 1918 | Miscellaneous |
| Molecular Sequence | 1453 | Basic Science and Research Tools |
| Antibiotic | 861 | Drugs and Chemicals |
| Drug Delivery Device | 790 | Drugs and Chemicals |
| Fully Formed Anatomical Structure | 104 | Clinical and Anatomy |
| Entity | 78 | Miscellaneous |
| Human | 44 | Miscellaneous |
| Vertebrate | 15 | Miscellaneous |

**Table S3: Examples of UMLS Terms.**

A UMLS term that is linked to multiple UMLS categories is treated as multiple separate observations; each such link represents one observation. All (UMLS term, UMLS category) pairs are first ranked based on the number of times the UMLS term is the newest term in a paper among all terms that belong to the same UMLS category.

We present 4 separate lists, one for each of the following four groups of idea categories that we use in the paper (Table S2 shows how the 127 UMLS categories map into these 4 category groups): “Clinical and Anatomy”, “Drugs and Chemicals”, “Basic Science and Research Tools”, and “Miscellaneous”

The rankings are constructed separately for each of these 4 idea category groups and for each decade, with the decade determined based on the cohort year of the UMLS term. The cohort year of a UMLS term is the year the term is first mentioned in the MEDLINE database. For each UMLS term the table also lists the earliest cohort of any of the term’s synonyms that appear in the UMLS metathesaurus.

For each decade we only present the top 25 UMLS terms. The analysis in the paper is based on all UMLS terms, not only the UMLS terms presented here. The focus on on a narrow set of popular UMLS terms here is for expositional convenience only.

Explanations for the columns:

Column (1): Decade of cohort; calculated based on the first number in column (6).
Column (2): Rank within decade of cohort; calculated based on column (3) and the first number in column (6).
Column (3): Number of times the UMLS term appears in a paper and is the newest term in the paper from that idea category. Calculated based on papers published during 2010-2016.
Column (4): Cumulative share of earliest mentions, calculated based on column (3) separately for each decade of cohort..
Column (5): The UMLS term.
Column (6): Cohort of term, set as the earliest year the term is mentioned in MEDLINE. The number in parenthesis is the earliest cohort of any synonym of the term (including the term itself).
Column (7): The UMLS category of the term; in our analysis this represents the idea category of the term.

The UMLS term lists for the 4 idea category groups appear in this order below: “Clinical and Anatomy”, “Drugs and Chemicals”, “Basic Science and Research Tools”, and “Miscellaneous”.

| (1) | (2) | (3) | (4) | (5) | (6) | (7) |
|---|---|---|---|---|---|---|
| CLINICAL AND ANATOMY (1st of 4 idea category groups) | | | | | | |
| 2010s | 1 | 780 | 1.98% | granulomatosis with polyangiitis | 2011 (1949) | Disease or Syndrome |
| 2010s | 2 | 698 | 3.76% | H7N9 | 2010 (1949) | Disease or Syndrome |
| 2010s | 3 | 388 | 4.75% | fecal microbiota transplantation | 2011 (2001) | Therapeutic or Preventive Procedure |
| 2010s | 4 | 365 | 5.68% | middle east respiratory syndrome | 2013 (1974) | Disease or Syndrome |
| 2010s | 5 | 279 | 6.39% | eosinophilic granulomatosis with polyangiitis | 2012 (2012) | Disease or Syndrome |
| 2010s | 6 | 182 | 6.85% | ecigarette user | 2011 (2010) | Finding |
| 2010s | 7 | 176 | 7.30% | H7N9 influenza | 2012 (2012) | Disease or Syndrome |
| 2010s | 8 | 150 | 7.68% | patientderived xenograft model | 2010 (1989) | Experimental Model of Disease |
| 2010s | 9 | 150 | 8.06% | vascularized composite allotransplantation | 2011 (1991) | Therapeutic or Preventive Procedure |
| 2010s | 10 | 146 | 8.43% | auditory neuropathy spectrum disorder | 2010 (1996) | Disease or Syndrome |
| 2010s | 11 | 143 | 8.80% | hoarding disorder | 2010 (2010) | Mental or Behavioral Dysfunction |
| 2010s | 12 | 132 | 9.13% | prostate health index | 2010 (1949) | Laboratory Procedure |
| 2010s | 13 | 125 | 9.45% | severe fever with thrombocytopenia syndrome | 2011 (2011) | Disease or Syndrome |
| 2010s | 14 | 117 | 9.75% | tedizolid | 2011 (2011) | Clinical Attribute |
| 2010s | 15 | 114 | 10.0% | C3 glomerulopathy | 2010 (2010) | Disease or Syndrome |
| 2010s | 16 | 112 | 10.3% | severe fever with thrombocytopenia syndrome virus | 2012 (2011) | Disease or Syndrome |
| 2010s | 17 | 108 | 10.6% | primary biliary cholangitis | 2015 (1949) | Disease or Syndrome |
| 2010s | 18 | 107 | 10.8% | florbetapir | 2010 (2010) | Indicator, Reagent, or Diagnostic Aid |
| 2010s | 19 | 102 | 11.1% | fusion biopsy | 2012 (2011) | Diagnostic Procedure |
| 2010s | 20 | 92 | 11.3% | mixed adenoneuroendocrine carcinoma | 2011 (1963) | Neoplastic Process |
| 2000s | 1 | 4562 | 1.88% | STEMI | 2000 (2000) | Finding |
| 2000s | 2 | 4516 | 3.75% | STEMI | 2000 (1994) | Disease or Syndrome |
| 2000s | 3 | 3811 | 5.33% | everolimus | 2000 (2000) | Laboratory Procedure |
| 2000s | 4 | 3292 | 6.69% | creactive protein hs | 2000 (2000) | Laboratory Procedure |
| 2000s | 5 | 3055 | 7.95% | castrationresistant prostate cancer | 2004 (1983) | Neoplastic Process |
| 2000s | 6 | 2977 | 9.18% | cardiac resynchronization therapy | 2000 (2000) | Therapeutic or Preventive Procedure |
| 2000s | 7 | 2928 | 10.3% | multidetector computed tomography | 2000 (1992) | Diagnostic Procedure |
| 2000s | 8 | 2888 | 11.5% | transcatheter aortic valve implantation | 2005 (1990) | Therapeutic or Preventive Procedure |
| 2000s | 9 | 2485 | 12.6% | positron emission tomography computed tomography | 2002 (1991) | Diagnostic Procedure |
| 2000s | 10 | 2313 | 13.5% | triplenegative breast cancer | 2006 (2006) | Finding |
| 2000s | 11 | 2131 | 14.4% | endoscopic submucosal dissection | 2004 (2004) | Therapeutic or Preventive Procedure |
| 2000s | 12 | 2041 | 15.3% | triplenegative breast cancer | 2006 (2006) | Neoplastic Process |
| 2000s | 13 | 1985 | 16.1% | CXCL10 | 2001 (1983) | Laboratory Procedure |
| 2000s | 14 | 1968 | 16.9% | transcranial direct current stimulation | 2000 (1987) | Therapeutic or Preventive Procedure |
| 2000s | 15 | 1618 | 17.6% | transcatheter aortic valve replacement | 2006 (1990) | Therapeutic or Preventive Procedure |
| 2000s | 16 | 1540 | 18.2% | transcriptome sequencing | 2007 (2007) | Laboratory Procedure |
| 2000s | 17 | 1455 | 18.8% | tigecycline | 2002 (2002) | Clinical Attribute |
| 2000s | 18 | 1443 | 19.4% | MELD score | 2001 (2001) | Clinical Attribute |
| 2000s | 19 | 1434 | 20.0% | MELD score | 2001 (2001) | Laboratory Procedure |
| 2000s | 20 | 1355 | 20.5% | takotsubo cardiomyopathy | 2000 (1976) | Disease or Syndrome |
| 1990s | 1 | 16494 | 2.08% | fmri | 1994 (1988) | Diagnostic Procedure |
| 1990s | 2 | 16180 | 4.12% | optical coherence tomography | 1991 (1991) | Diagnostic Procedure |
| 1990s | 3 | 12851 | 5.75% | percutaneous coronary intervention | 1991 (1991) | Therapeutic or Preventive Procedure |
| 1990s | 4 | 9244 | 6.92% | adiponectin | 1999 (1999) | Laboratory Procedure |
| 1990s | 5 | 8538 | 7.99% | microarray analysis | 1998 (1989) | Laboratory Procedure |
| 1990s | 6 | 7631 | 8.96% | chromatin immunoprecipitation | 1998 (1949) | Laboratory Procedure |
| 1990s | 7 | 6933 | 9.83% | MMP9 | 1991 (1991) | Laboratory Procedure |
| 1990s | 8 | 6657 | 10.6% | pyrosequencing | 1998 (1998) | Laboratory Procedure |
| 1990s | 9 | 6447 | 11.4% | autism spectrum disorder | 1992 (1992) | Finding |
| 1990s | 10 | 6188 | 12.2% | diffusion tensor imaging | 1994 (1994) | Diagnostic Procedure |
| 1990s | 11 | 5886 | 13.0% | NAFLD | 1998 (1977) | Disease or Syndrome |
| 1990s | 12 | 5528 | 13.7% | autism spectrum disorder | 1992 (1981) | Mental or Behavioral Dysfunction |
| 1990s | 13 | 5398 | 14.4% | gene expression profiling | 1998 (1989) | Laboratory Procedure |
| 1990s | 14 | 4795 | 15.0% | autism spectrum disorders | 1992 (1982) | Mental or Behavioral Dysfunction |
| 1990s | 15 | 4314 | 15.5% | tacrolimus | 1992 (1992) | Laboratory Procedure |
| 1990s | 16 | 4295 | 16.0% | BRCA1 | 1993 (1993) | Laboratory Procedure |
| 1990s | 17 | 3722 | 16.5% | ghrelin | 1999 (1989) | Laboratory Procedure |
| 1990s | 18 | 3535 | 17.0% | highly active antiretroviral therapy | 1996 (1970) | Therapeutic or Preventive Procedure |
| 1990s | 19 | 3534 | 17.4% | microcomputed tomography | 1990 (1975) | Diagnostic Procedure |
| 1990s | 20 | 3138 | 17.8% | statin therapy | 1993 (1993) | Therapeutic or Preventive Procedure |
| 1980s | 1 | 30408 | 1.53% | polymerase chain reaction | 1986 (1986) | Laboratory Procedure |
| 1980s | 2 | 19973 | 2.53% | western blot | 1981 (1981) | Laboratory Procedure |
| 1980s | 3 | 18078 | 3.45% | primary endpoint | 1980 (1980) | Indicator, Reagent, or Diagnostic Aid |
| 1980s | 4 | 17719 | 4.34% | HIV1 | 1986 (1986) | Laboratory or Test Result |
| 1980s | 5 | 17599 | 5.23% | VEGF | 1987 (1982) | Laboratory Procedure |
| 1980s | 6 | 17530 | 6.11% | vascular endothelial growth factor | 1982 (1982) | Therapeutic or Preventive Procedure |
| 1980s | 7 | 15337 | 6.88% | tissue engineering | 1984 (1984) | Therapeutic or Preventive Procedure |
| 1980s | 8 | 15075 | 7.64% | NSCLC | 1981 (1976) | Neoplastic Process |
| 1980s | 9 | 14097 | 8.35% | western blot analysis | 1982 (1981) | Laboratory Procedure |
| 1980s | 10 | 13523 | 9.04% | HIV infection | 1986 (1986) | Clinical Attribute |
| 1980s | 11 | 12977 | 9.69% | neuroimaging | 1982 (1982) | Diagnostic Procedure |
| 1980s | 12 | 12828 | 10.3% | antiretroviral therapy | 1985 (1985) | Therapeutic or Preventive Procedure |
| 1980s | 13 | 11657 | 10.9% | EGFR | 1980 (1977) | Laboratory Procedure |
| 1980s | 14 | 11549 | 11.5% | atomic force microscopy | 1988 (1976) | Laboratory Procedure |
| 1980s | 15 | 11367 | 12.0% | LCMS | 1982 (1970) | Laboratory Procedure |
| 1980s | 16 | 10204 | 12.5% | HIV infection | 1986 (1983) | Disease or Syndrome |
| 1980s | 17 | 9567 | 13.0% | human immunodeficiency virus | 1986 (1983) | Disease or Syndrome |
| 1980s | 18 | 9546 | 13.5% | confocal microscopy | 1981 (1981) | Laboratory Procedure |
| 1980s | 19 | 8799 | 14.0% | interleukin6 | 1987 (1987) | Laboratory Procedure |
| 1980s | 20 | 8322 | 14.4% | PTSD | 1982 (1949) | Mental or Behavioral Dysfunction |
| 1970s | 1 | 71568 | 2.12% | biomarkers | 1973 (1949) | Clinical Attribute |
| 1970s | 2 | 44237 | 3.43% | magnetic resonance imaging | 1978 (1949) | Diagnostic Procedure |
| 1970s | 3 | 40954 | 4.64% | body mass index | 1975 (1975) | Diagnostic Procedure |
| 1970s | 4 | 35912 | 5.71% | biomarker | 1973 (1949) | Clinical Attribute |
| 1970s | 5 | 34725 | 6.74% | body mass index | 1975 (1970) | Clinical Attribute |
| 1970s | 6 | 34105 | 7.75% | body mass index | 1975 (1975) | Finding |
| 1970s | 7 | 27597 | 8.56% | body mass index BMI | 1978 (1978) | Clinical Attribute |
| 1970s | 8 | 23495 | 9.26% | flow cytometry | 1977 (1971) | Laboratory Procedure |
| 1970s | 9 | 18772 | 9.82% | treatment options | 1971 (1950) | Therapeutic or Preventive Procedure |
| 1970s | 10 | 17892 | 10.3% | T cells | 1970 (1970) | Laboratory Procedure |
| 1970s | 11 | 17877 | 10.8% | HPLC | 1973 (1969) | Laboratory Procedure |
| 1970s | 12 | 17627 | 11.4% | risk assessment | 1973 (1973) | Health Care Activity |
| 1970s | 13 | 15331 | 11.8% | ELISA | 1971 (1971) | Laboratory Procedure |
| 1970s | 14 | 13416 | 12.2% | CD8 | 1979 (1979) | Laboratory Procedure |
| 1970s | 15 | 12428 | 12.6% | interventional | 1971 (1971) | Diagnostic Procedure |
| 1970s | 16 | 11965 | 12.9% | neurodegeneration | 1976 (1976) | Finding |
| 1970s | 17 | 11292 | 13.3% | cancer progression | 1979 (1979) | Pathologic Function |
| 1970s | 18 | 11170 | 13.6% | neurodegenerative diseases | 1979 (1965) | Disease or Syndrome |
| 1970s | 19 | 10573 | 13.9% | poor outcome | 1975 (1975) | Finding |
| 1970s | 20 | 10449 | 14.2% | working memory | 1977 (1949) | Mental Process |
| 1960s | 1 | 59518 | 2.01% | immunohistochemistry | 1964 (1964) | Diagnostic Procedure |
| 1960s | 2 | 48580 | 3.65% | mouse model | 1965 (1965) | Experimental Model of Disease |
| 1960s | 3 | 38862 | 4.96% | sequencing | 1962 (1962) | Laboratory Procedure |
| 1960s | 4 | 24939 | 5.81% | scanning electron microscopy | 1963 (1963) | Diagnostic Procedure |
| 1960s | 5 | 24752 | 6.64% | colorectal cancer | 1962 (1962) | Finding |
| 1960s | 6 | 23258 | 7.43% | colorectal cancer | 1962 (1949) | Neoplastic Process |
| 1960s | 7 | 19812 | 8.10% | ethnicity | 1966 (1966) | Clinical Attribute |
| 1960s | 8 | 15572 | 8.62% | ethnicity | 1966 (1966) | Finding |
| 1960s | 9 | 14815 | 9.13% | crosstalk | 1966 (1966) | Injury or Poisoning |
| 1960s | 10 | 13656 | 9.59% | scanning electron microscopy | 1963 (1963) | Laboratory Procedure |
| 1960s | 11 | 13069 | 10.0% | COPD | 1967 (1949) | Disease or Syndrome |
| 1960s | 12 | 12966 | 10.4% | ischemic stroke | 1963 (1963) | Finding |
| 1960s | 13 | 12852 | 10.9% | coherent | 1961 (1961) | Finding |
| 1960s | 14 | 12795 | 11.3% | immunosuppression | 1964 (1964) | Pathologic Function |
| 1960s | 15 | 12684 | 11.7% | transmission electron microscopy | 1964 (1949) | Laboratory Procedure |
| 1960s | 16 | 12680 | 12.1% | chart review | 1966 (1957) | Health Care Activity |
| 1960s | 17 | 12659 | 12.6% | ischemic stroke | 1963 (1962) | Disease or Syndrome |
| 1960s | 18 | 12121 | 13.0% | high risk of | 1961 (1955) | Finding |
| 1960s | 19 | 11154 | 13.4% | NMR spectroscopy | 1965 (1961) | Diagnostic Procedure |
| 1960s | 20 | 11087 | 13.7% | inflammatory bowel disease | 1964 (1964) | Finding |
| 1950s | 1 | 222028 | 4.98% | strategies | 1955 (1955) | Educational Activity |
| 1950s | 2 | 213336 | 9.77% | strategies | 1955 (1949) | Mental Process |
| 1950s | 3 | 75034 | 11.4% | quality of life | 1959 (1959) | Sign or Symptom |
| 1950s | 4 | 69350 | 13.0% | risk factors | 1959 (1959) | Finding |
| 1950s | 5 | 58653 | 14.3% | encoding | 1956 (1953) | Mental Process |
| 1950s | 6 | 56850 | 15.6% | documented | 1950 (1950) | Health Care Activity |
| 1950s | 7 | 45073 | 16.6% | quality of life | 1959 (1959) | Finding |
| 1950s | 8 | 25692 | 17.1% | options | 1950 (1950) | Therapeutic or Preventive Procedure |
| 1950s | 9 | 24733 | 17.7% | risk factor | 1959 (1959) | Finding |
| 1950s | 10 | 24543 | 18.3% | pharmacokinetics | 1955 (1949) | Physiologic Function |
| 1950s | 11 | 23834 | 18.8% | immune responses | 1950 (1949) | Organ or Tissue Function |
| 1950s | 12 | 23790 | 19.3% | immunohistochemical | 1956 (1956) | Laboratory Procedure |
| 1950s | 13 | 23673 | 19.9% | encoded | 1953 (1953) | Mental Process |
| 1950s | 14 | 22955 | 20.4% | hepatocellular carcinoma | 1951 (1949) | Neoplastic Process |
| 1950s | 15 | 22021 | 20.9% | computed tomography | 1956 (1949) | Diagnostic Procedure |
| 1950s | 16 | 21976 | 21.4% | high risk | 1955 (1955) | Health Care Activity |
| 1950s | 17 | 21905 | 21.9% | hepatocellular carcinoma | 1951 (1951) | Finding |
| 1950s | 18 | 20433 | 22.3% | triggers | 1955 (1949) | Clinical Attribute |
| 1950s | 19 | 19918 | 22.8% | animal model | 1954 (1954) | Experimental Model of Disease |
| 1950s | 20 | 19524 | 23.2% | laparoscopic | 1950 (1949) | Diagnostic Procedure |
| | | | | | | |
| | | | | | | |
| DRUGS AND CHEMICALS (2nd of 4 idea category groups) | | | | | | |
| 2010s | 1 | 856 | 1.29% | crizotinib | 2010 (2010) | Pharmacologic Substance |
| 2010s | 2 | 851 | 2.57% | vemurafenib | 2011 (2011) | Pharmacologic Substance |
| 2010s | 3 | 686 | 3.61% | enzalutamide | 2012 (2012) | Pharmacologic Substance |
| 2010s | 4 | 465 | 4.31% | ibrutinib | 2012 (2012) | Pharmacologic Substance |
| 2010s | 5 | 457 | 5.00% | ruxolitinib | 2010 (2010) | Pharmacologic Substance |
| 2010s | 6 | 449 | 5.68% | nivolumab | 2013 (2013) | Pharmacologic Substance |
| 2010s | 7 | 438 | 6.34% | afatinib | 2011 (2011) | Pharmacologic Substance |
| 2010s | 8 | 433 | 7.00% | pembrolizumab | 2014 (2013) | Pharmacologic Substance |
| 2010s | 9 | 410 | 7.61% | sofosbuvir | 2013 (2013) | Pharmacologic Substance |
| 2010s | 10 | 384 | 8.19% | dabrafenib | 2012 (2012) | Pharmacologic Substance |
| 2010s | 11 | 336 | 8.70% | simeprevir | 2013 (2008) | Pharmacologic Substance |
| 2010s | 12 | 329 | 9.20% | tofacitinib | 2010 (2008) | Pharmacologic Substance |
| 2010s | 13 | 326 | 9.69% | regorafenib | 2011 (2011) | Pharmacologic Substance |
| 2010s | 14 | 318 | 10.1% | brentuximab vedotin | 2010 (2003) | Pharmacologic Substance |
| 2010s | 15 | 311 | 10.6% | dolutegravir | 2011 (2011) | Pharmacologic Substance |
| 2010s | 16 | 308 | 11.1% | empagliflozin | 2012 (2012) | Pharmacologic Substance |
| 2010s | 17 | 268 | 11.5% | canagliflozin | 2010 (2010) | Pharmacologic Substance |
| 2010s | 18 | 256 | 11.9% | vismodegib | 2010 (2010) | Pharmacologic Substance |
| 2010s | 19 | 251 | 12.2% | ponatinib | 2011 (2011) | Pharmacologic Substance |
| 2010s | 20 | 230 | 12.6% | nintedanib | 2012 (2010) | Pharmacologic Substance |
| 2000s | 1 | 6248 | 2.61% | bevacizumab | 2001 (1992) | Pharmacologic Substance |
| 2000s | 2 | 3130 | 3.93% | sorafenib | 2004 (2004) | Pharmacologic Substance |
| 2000s | 3 | 3016 | 5.19% | imatinib | 2001 (2001) | Pharmacologic Substance |
| 2000s | 4 | 2694 | 6.32% | bortezomib | 2002 (2002) | Pharmacologic Substance |
| 2000s | 5 | 2646 | 7.43% | everolimus | 2000 (1997) | Pharmacologic Substance |
| 2000s | 6 | 2501 | 8.48% | sunitinib | 2005 (2005) | Pharmacologic Substance |
| 2000s | 7 | 2377 | 9.47% | erlotinib | 2002 (2002) | Pharmacologic Substance |
| 2000s | 8 | 2290 | 10.4% | adalimumab | 2002 (2002) | Pharmacologic Substance |
| 2000s | 9 | 1993 | 11.2% | cetuximab | 2000 (1984) | Pharmacologic Substance |
| 2000s | 10 | 1932 | 12.0% | CXCL10 | 2001 (1974) | Pharmacologic Substance |
| 2000s | 11 | 1844 | 12.8% | rivaroxaban | 2006 (2005) | Pharmacologic Substance |
| 2000s | 12 | 1801 | 13.6% | ranibizumab | 2003 (2003) | Pharmacologic Substance |
| 2000s | 13 | 1788 | 14.3% | lenalidomide | 2004 (2001) | Pharmacologic Substance |
| 2000s | 14 | 1771 | 15.1% | zoledronic acid | 2000 (2000) | Clinical Drug |
| 2000s | 15 | 1527 | 15.7% | rosuvastatin | 2001 (2001) | Pharmacologic Substance |
| 2000s | 16 | 1471 | 16.3% | gefitinib | 2002 (2002) | Pharmacologic Substance |
| 2000s | 17 | 1469 | 16.9% | SP600125 | 2001 (2001) | Pharmacologic Substance |
| 2000s | 18 | 1454 | 17.5% | tigecycline | 2002 (1999) | Organic Chemical |
| 2000s | 19 | 1371 | 18.1% | zoledronic acid | 2000 (2000) | Pharmacologic Substance |
| 2000s | 20 | 1200 | 18.6% | denosumab | 2005 (2005) | Pharmacologic Substance |
| 1990s | 1 | 16209 | 3.94% | IL10 | 1990 (1990) | Pharmacologic Substance |
| 1990s | 2 | 12716 | 7.03% | antiapoptotic | 1992 (1992) | Chemical Viewed Functionally |
| 1990s | 3 | 11306 | 9.78% | carbon nanotubes | 1992 (1969) | Chemical Viewed Structurally |
| 1990s | 4 | 7256 | 11.5% | rituximab | 1997 (1987) | Pharmacologic Substance |
| 1990s | 5 | 6735 | 13.1% | paclitaxel | 1993 (1993) | Pharmacologic Substance |
| 1990s | 6 | 3940 | 14.1% | IL13 | 1993 (1992) | Pharmacologic Substance |
| 1990s | 7 | 3747 | 15.0% | clopidogrel | 1991 (1991) | Pharmacologic Substance |
| 1990s | 8 | 3408 | 15.8% | gemcitabine | 1990 (1985) | Pharmacologic Substance |
| 1990s | 9 | 3402 | 16.7% | docetaxel | 1993 (1993) | Pharmacologic Substance |
| 1990s | 10 | 2994 | 17.4% | trastuzumab | 1998 (1990) | Pharmacologic Substance |
| 1990s | 11 | 2937 | 18.1% | carbon nanotube | 1992 (1969) | Chemical Viewed Structurally |
| 1990s | 12 | 2738 | 18.8% | sirolimus | 1994 (1975) | Organic Chemical |
| 1990s | 13 | 2699 | 19.4% | infliximab | 1998 (1958) | Pharmacologic Substance |
| 1990s | 14 | 2687 | 20.1% | biodiesel | 1994 (1994) | Organic Chemical |
| 1990s | 15 | 2651 | 20.7% | tacrolimus | 1992 (1991) | Pharmacologic Substance |
| 1990s | 16 | 2296 | 21.3% | atorvastatin | 1994 (1994) | Pharmacologic Substance |
| 1990s | 17 | 2210 | 21.8% | linezolid | 1997 (1997) | Pharmacologic Substance |
| 1990s | 18 | 2138 | 22.3% | dendrimers | 1990 (1980) | Biomedical or Dental Material |
| 1990s | 19 | 2104 | 22.9% | endocannabinoid | 1997 (1991) | Biologically Active Substance |
| 1990s | 20 | 1940 | 23.3% | LY294002 | 1994 (1994) | Pharmacologic Substance |
| 1980s | 1 | 14480 | 2.61% | IL6 | 1987 (1982) | Pharmacologic Substance |
| 1980s | 2 | 13347 | 5.02% | VEGF | 1987 (1952) | Pharmacologic Substance |
| 1980s | 3 | 9907 | 6.81% | signaling molecule | 1982 (1982) | Biologically Active Substance |
| 1980s | 4 | 8898 | 8.42% | interleukin6 | 1987 (1982) | Pharmacologic Substance |
| 1980s | 5 | 6787 | 9.64% | HER2 | 1987 (1987) | Pharmacologic Substance |
| 1980s | 6 | 6216 | 10.7% | statins | 1983 (1949) | Pharmacologic Substance |
| 1980s | 7 | 6056 | 11.8% | IL8 | 1989 (1969) | Pharmacologic Substance |
| 1980s | 8 | 5397 | 12.8% | 3UTR | 1984 (1984) | Biologically Active Substance |
| 1980s | 9 | 4941 | 13.7% | oxaliplatin | 1989 (1989) | Clinical Drug |
| 1980s | 10 | 4814 | 14.5% | brainderived neurotrophic factor | 1985 (1985) | Pharmacologic Substance |
| 1980s | 11 | 4691 | 15.4% | ciprofloxacin | 1983 (1983) | Pharmacologic Substance |
| 1980s | 12 | 4630 | 16.2% | ciprofloxacin | 1983 (1983) | Organic Chemical |
| 1980s | 13 | 4199 | 17.0% | IGF1 | 1980 (1974) | Pharmacologic Substance |
| 1980s | 14 | 3791 | 17.7% | propofol | 1984 (1980) | Pharmacologic Substance |
| 1980s | 15 | 3571 | 18.3% | vascular endothelial growth factor | 1982 (1952) | Pharmacologic Substance |
| 1980s | 16 | 3538 | 19.0% | interleukin | 1980 (1980) | Pharmacologic Substance |
| 1980s | 17 | 3535 | 19.6% | protein kinase C | 1981 (1981) | Pharmacologic Substance |
| 1980s | 18 | 3490 | 20.2% | interleukin1 | 1980 (1970) | Pharmacologic Substance |
| 1980s | 19 | 3433 | 20.8% | fluconazole | 1985 (1985) | Clinical Drug |
| 1980s | 20 | 3298 | 21.4% | temozolomide | 1988 (1988) | Clinical Drug |
| 1970s | 1 | 13252 | 2.36% | monoclonal antibodies | 1971 (1971) | Clinical Drug |
| 1970s | 2 | 10750 | 4.28% | doxorubicin | 1972 (1949) | Organic Chemical |
| 1970s | 3 | 8736 | 5.84% | logran | 1973 (1973) | Pharmacologic Substance |
| 1970s | 4 | 8044 | 7.27% | intron | 1978 (1978) | Pharmacologic Substance |
| 1970s | 5 | 7741 | 8.65% | monoclonal antibodies | 1971 (1971) | Pharmacologic Substance |
| 1970s | 6 | 7047 | 9.91% | cisplatin | 1971 (1970) | Pharmacologic Substance |
| 1970s | 7 | 6346 | 11.0% | immunomodulator | 1976 (1949) | Pharmacologic Substance |
| 1970s | 8 | 5985 | 12.1% | peroxisome proliferator | 1975 (1975) | Hazardous or Poisonous Substance |
| 1970s | 9 | 5841 | 13.1% | tumor necrosis factor | 1975 (1975) | Pharmacologic Substance |
| 1970s | 10 | 5465 | 14.1% | rapamycin | 1975 (1975) | Organic Chemical |
| 1970s | 11 | 4992 | 15.0% | 25hydroxyvitamin D | 1973 (1973) | Vitamin |
| 1970s | 12 | 4797 | 15.8% | pristine | 1974 (1974) | Hazardous or Poisonous Substance |
| 1970s | 13 | 4339 | 16.6% | IL1 | 1976 (1970) | Pharmacologic Substance |
| 1970s | 14 | 4319 | 17.4% | pristine | 1974 (1974) | Organic Chemical |
| 1970s | 15 | 4179 | 18.1% | 25hydroxyvitamin D | 1973 (1968) | Pharmacologic Substance |
| 1970s | 16 | 4053 | 18.8% | antimicrobial peptide | 1979 (1979) | Pharmacologic Substance |
| 1970s | 17 | 3232 | 19.4% | resveratrol | 1978 (1978) | Pharmacologic Substance |
| 1970s | 18 | 3122 | 20.0% | LDL cholesterol | 1972 (1955) | Biologically Active Substance |
| 1970s | 19 | 2682 | 20.5% | angiogenic factor | 1973 (1972) | Biologically Active Substance |
| 1970s | 20 | 2639 | 20.9% | tumor markers | 1973 (1973) | Biologically Active Substance |
| 1960s | 1 | 52920 | 7.17% | ligands | 1960 (1949) | Chemical |
| 1960s | 2 | 15286 | 9.24% | superoxide dismutase | 1969 (1969) | Organic Chemical |
| 1960s | 3 | 13413 | 11.0% | COPD | 1967 (1967) | Pharmacologic Substance |
| 1960s | 4 | 10744 | 12.5% | superoxide dismutase | 1969 (1949) | Pharmacologic Substance |
| 1960s | 5 | 9858 | 13.8% | molecular target | 1969 (1969) | Chemical Viewed Functionally |
| 1960s | 6 | 9517 | 15.1% | xenografts | 1962 (1949) | Biomedical or Dental Material |
| 1960s | 7 | 9335 | 16.4% | xenograft | 1962 (1949) | Biomedical or Dental Material |
| 1960s | 8 | 9068 | 17.6% | opioids | 1968 (1968) | Biologically Active Substance |
| 1960s | 9 | 8209 | 18.7% | allograft | 1963 (1949) | Biomedical or Dental Material |
| 1960s | 10 | 7872 | 19.8% | biomaterials | 1967 (1967) | Biomedical or Dental Material |
| 1960s | 11 | 6333 | 20.6% | bioi | 1960 (1960) | Inorganic Chemical |
| 1960s | 12 | 5967 | 21.4% | dopaminergic | 1964 (1964) | Pharmacologic Substance |
| 1960s | 13 | 5773 | 22.2% | hotspot | 1961 (1961) | Pharmacologic Substance |
| 1960s | 14 | 5431 | 22.9% | CI 4 | 1960 (1960) | Pharmacologic Substance |
| 1960s | 15 | 5414 | 23.7% | hydrogels | 1964 (1964) | Biomedical or Dental Material |
| 1960s | 16 | 5038 | 24.4% | pahs | 1965 (1949) | Organic Chemical |
| 1960s | 17 | 4882 | 25.0% | immunosuppressive | 1963 (1963) | Pharmacologic Substance |
| 1960s | 18 | 4712 | 25.7% | neurotransmitters | 1962 (1955) | Biologically Active Substance |
| 1960s | 19 | 4677 | 26.3% | opioids | 1968 (1949) | Pharmacologic Substance |
| 1960s | 20 | 3803 | 26.8% | allografts | 1963 (1949) | Biomedical or Dental Material |
| 1950s | 1 | 17863 | 2.36% | remodelin | 1950 (1950) | Pharmacologic Substance |
| 1950s | 2 | 17123 | 4.63% | oncogen | 1950 (1949) | Hazardous or Poisonous Substance |
| 1950s | 3 | 15506 | 6.68% | lipopolysaccharide | 1950 (1950) | Organic Chemical |
| 1950s | 4 | 13158 | 8.42% | malondialdehyde | 1951 (1951) | Biologically Active Substance |
| 1950s | 5 | 11963 | 10.0% | mimics | 1950 (1949) | Hazardous or Poisonous Substance |
| 1950s | 6 | 11734 | 11.5% | surfactant | 1951 (1951) | Biologically Active Substance |
| 1950s | 7 | 11396 | 13.0% | arabidopsis thaliana | 1955 (1955) | Organic Chemical |
| 1950s | 8 | 10802 | 14.5% | surfactant | 1951 (1949) | Biomedical or Dental Material |
| 1950s | 9 | 8975 | 15.6% | surfactant | 1951 (1951) | Chemical Viewed Functionally |
| 1950s | 10 | 8534 | 16.8% | pollutants | 1950 (1950) | Hazardous or Poisonous Substance |
| 1950s | 11 | 7060 | 17.7% | predef | 1950 (1950) | Pharmacologic Substance |
| 1950s | 12 | 7048 | 18.6% | streptozotocin | 1959 (1959) | Organic Chemical |
| 1950s | 13 | 6844 | 19.5% | hydrogel | 1955 (1955) | Pharmacologic Substance |
| 1950s | 14 | 6840 | 20.4% | cortisol | 1954 (1949) | Pharmacologic Substance |
| 1950s | 15 | 6690 | 21.3% | interferon | 1957 (1957) | Pharmacologic Substance |
| 1950s | 16 | 6588 | 22.2% | virulence factors | 1952 (1949) | Hazardous or Poisonous Substance |
| 1950s | 17 | 6406 | 23.1% | dopamine | 1952 (1949) | Pharmacologic Substance |
| 1950s | 18 | 6239 | 23.9% | neurotransmitter | 1955 (1955) | Biologically Active Substance |
| 1950s | 19 | 5924 | 24.7% | surfactants | 1951 (1951) | Chemical Viewed Functionally |
| 1950s | 20 | 5866 | 25.4% | agonist | 1952 (1952) | Pharmacologic Substance |

| BASIC SCIENCE AND RESEARCH TOOLS (3rd of 4 idea category groups) | | | | | | |
|---|---|---|---|---|---|---|
| 2010s | 1 | 663 | .666% | mechanistic target of rapamycin | 2010 (1971) | Amino Acid, Peptide, or Protein |
| 2010s | 2 | 658 | 1.32% | mechanistic target of rapamycin | 2010 (1976) | Gene or Genome |
| 2010s | 3 | 648 | 1.97% | middle east respiratory syndrome coronavirus | 2013 (1999) | Virus |
| 2010s | 4 | 403 | 2.38% | transcription activatorlike effector nucleases | 2011 (2010) | Amino Acid, Peptide, or Protein |
| 2010s | 5 | 323 | 2.70% | AMPK1 | 2010 (1985) | Gene or Genome |
| 2010s | 6 | 304 | 3.01% | H7N9 virus | 2013 (2013) | Virus |
| 2010s | 7 | 301 | 3.31% | C9ORF72 | 2011 (2011) | Gene or Genome |
| 2010s | 8 | 295 | 3.61% | schmallenberg virus | 2012 (2012) | Virus |
| 2010s | 9 | 276 | 3.88% | talens | 2010 (2010) | Amino Acid, Peptide, or Protein |
| 2010s | 10 | 256 | 4.14% | interleukin28b | 2010 (2003) | Amino Acid, Peptide, or Protein |
| 2010s | 11 | 246 | 4.39% | crisprcas systems | 2011 (2006) | Molecular Function |
| 2010s | 12 | 219 | 4.61% | mechanistic target of rapamycin complex 1 | 2010 (2002) | Amino Acid, Peptide, or Protein |
| 2010s | 13 | 214 | 4.82% | telocytes | 2010 (2005) | Cell |
| 2010s | 14 | 191 | 5.02% | SRSF2 | 2011 (1991) | Amino Acid, Peptide, or Protein |
| 2010s | 15 | 182 | 5.20% | chromothripsis | 2011 (1954) | Cell or Molecular Dysfunction |
| 2010s | 16 | 179 | 5.38% | CALR mutation | 2013 (2013) | Cell or Molecular Dysfunction |
| 2010s | 17 | 164 | 5.54% | fukushima nuclear accident | 2011 (2011) | Human-caused Phenomenon or Process |
| 2010s | 18 | 162 | 5.71% | beige adipocytes | 2012 (2010) | Cell |
| 2010s | 19 | 151 | 5.86% | SRSF2 | 2011 (1991) | Gene or Genome |
| 2010s | 20 | 150 | 6.01% | ocriplasmin | 2010 (1987) | Amino Acid, Peptide, or Protein |
| 2000s | 1 | 21210 | 3.08% | micrornas | 2001 (1971) | Nucleic Acid, Nucleoside, or Nucleotide |
| 2000s | 2 | 12390 | 4.88% | microrna | 2000 (1971) | Nucleic Acid, Nucleoside, or Nucleotide |
| 2000s | 3 | 9139 | 6.21% | nextgeneration sequencing | 2007 (2005) | Molecular Biology Research Technique |
| 2000s | 4 | 8606 | 7.46% | small interfering RNA | 2001 (1949) | Nucleic Acid, Nucleoside, or Nucleotide |
| 2000s | 5 | 6569 | 8.42% | GWAS | 2007 (1982) | Molecular Biology Research Technique |
| 2000s | 6 | 4290 | 9.04% | induced pluripotent stem cells | 2006 (1966) | Cell |
| 2000s | 7 | 3885 | 9.61% | th17 cells | 2006 (1980) | Cell |
| 2000s | 8 | 3405 | 10.1% | deep sequencing | 2000 (2000) | Molecular Biology Research Technique |
| 2000s | 9 | 3055 | 10.5% | mtorc1 | 2004 (2002) | Cell Component |
| 2000s | 10 | 2976 | 10.9% | IL17A | 2003 (1988) | Amino Acid, Peptide, or Protein |
| 2000s | 11 | 2746 | 11.3% | IL17A | 2003 (1988) | Gene or Genome |
| 2000s | 12 | 2667 | 11.7% | CD133 | 2000 (2000) | Amino Acid, Peptide, or Protein |
| 2000s | 13 | 2651 | 12.1% | inflammasome | 2002 (2002) | Amino Acid, Peptide, or Protein |
| 2000s | 14 | 2532 | 12.5% | short hairpin RNA | 2002 (1982) | Nucleic Acid, Nucleoside, or Nucleotide |
| 2000s | 15 | 2520 | 12.8% | exome sequencing | 2009 (2009) | Molecular Biology Research Technique |
| 2000s | 16 | 2513 | 13.2% | CD133 | 2000 (1978) | Gene or Genome |
| 2000s | 17 | 2466 | 13.6% | norovirus | 2002 (2002) | Amino Acid, Peptide, or Protein |
| 2000s | 18 | 2411 | 13.9% | small interfering rna | 2001 (1949) | Nucleic Acid, Nucleoside, or Nucleotide |
| 2000s | 19 | 2318 | 14.3% | IL23 | 2000 (2000) | Molecular Function |
| 2000s | 20 | 2265 | 14.6% | long noncoding RNA | 2007 (2003) | Nucleic Acid, Nucleoside, or Nucleotide |
| 1990s | 1 | 20439 | 1.19% | graphene | 1992 (1992) | Element, Ion, or Isotope |
| 1990s | 2 | 20423 | 2.39% | realtime PCR | 1996 (1989) | Molecular Biology Research Technique |
| 1990s | 3 | 18527 | 3.47% | single nucleotide polymorphisms | 1994 (1966) | Nucleotide Sequence |
| 1990s | 4 | 17023 | 4.46% | IL10 | 1990 (1990) | Molecular Function |
| 1990s | 5 | 14482 | 5.31% | transcriptome | 1997 (1997) | Nucleotide Sequence |
| 1990s | 6 | 14459 | 6.16% | caspase3 | 1997 (1949) | Gene or Genome |
| 1990s | 7 | 13862 | 6.97% | caspase3 | 1997 (1949) | Amino Acid, Peptide, or Protein |
| 1990s | 8 | 11694 | 7.65% | PI3K | 1990 (1989) | Molecular Function |
| 1990s | 9 | 10248 | 8.25% | nanomaterials | 1994 (1994) | Research Activity |
| 1990s | 10 | 9951 | 8.83% | qrtpcr | 1997 (1992) | Molecular Biology Research Technique |
| 1990s | 11 | 9237 | 9.37% | IL10 | 1990 (1975) | Gene or Genome |
| 1990s | 12 | 8821 | 9.89% | MAPK | 1990 (1959) | Molecular Function |
| 1990s | 13 | 8771 | 10.4% | quantitative realtime PCR | 1999 (1989) | Molecular Biology Research Technique |
| 1990s | 14 | 8459 | 10.9% | MMP9 | 1991 (1991) | Molecular Function |
| 1990s | 15 | 8432 | 11.3% | IL10 | 1990 (1975) | Amino Acid, Peptide, or Protein |
| 1990s | 16 | 8158 | 11.8% | adiponectin | 1999 (1966) | Gene or Genome |
| 1990s | 17 | 7965 | 12.3% | realtime polymerase chain reaction | 1997 (1989) | Molecular Biology Research Technique |
| 1990s | 18 | 7533 | 12.7% | adiponectin | 1999 (1982) | Amino Acid, Peptide, or Protein |
| 1990s | 19 | 7429 | 13.2% | single nucleotide polymorphism | 1991 (1966) | Nucleotide Sequence |
| 1990s | 20 | 6443 | 13.5% | PI3K | 1990 (1975) | Gene or Genome |
| 1980s | 1 | 52779 | 2.85% | signaling pathway | 1984 (1949) | Cell Function |
| 1980s | 2 | 41362 | 5.10% | signaling pathway | 1984 (1984) | Molecular Function |
| 1980s | 3 | 31713 | 6.81% | polymerase chain reaction | 1986 (1986) | Molecular Biology Research Technique |
| 1980s | 4 | 29573 | 8.42% | RTPCR | 1989 (1989) | Molecular Biology Research Technique |
| 1980s | 5 | 22089 | 9.61% | IL6 | 1987 (1987) | Molecular Function |
| 1980s | 6 | 17648 | 10.5% | western blotting | 1981 (1980) | Molecular Biology Research Technique |
| 1980s | 7 | 16542 | 11.4% | western blot | 1981 (1980) | Molecular Biology Research Technique |
| 1980s | 8 | 14755 | 12.2% | metaanalyses | 1982 (1975) | Research Activity |
| 1980s | 9 | 13893 | 13.0% | MTT assay | 1985 (1985) | Research Activity |
| 1980s | 10 | 13784 | 13.7% | HIV1 | 1986 (1984) | Virus |
| 1980s | 11 | 13393 | 14.4% | vascular endothelial growth factor | 1982 (1982) | Molecular Function |
| 1980s | 12 | 12584 | 15.1% | bcl2 | 1984 (1984) | Molecular Function |
| 1980s | 13 | 12158 | 15.8% | tandem mass spectrometry | 1981 (1952) | Molecular Biology Research Technique |
| 1980s | 14 | 11356 | 16.4% | RNA interference | 1987 (1959) | Genetic Function |
| 1980s | 15 | 10633 | 17.0% | EGFR | 1980 (1979) | Molecular Function |
| 1980s | 16 | 10614 | 17.6% | quantitative PCR | 1989 (1989) | Molecular Biology Research Technique |
| 1980s | 17 | 10205 | 18.1% | human immunodeficiency virus | 1986 (1983) | Virus |
| 1980s | 18 | 9966 | 18.6% | hepatitis C virus HCV | 1989 (1961) | Virus |
| 1980s | 19 | 8766 | 19.1% | mscs | 1980 (1980) | Molecular Function |
| 1980s | 20 | 8766 | 19.6% | VEGF | 1987 (1987) | Gene or Genome |
| 1970s | 1 | 88250 | 3.17% | targeting | 1971 (1969) | Cell Function |
| 1970s | 2 | 71032 | 5.73% | apoptosis | 1972 (1965) | Cell Function |
| 1970s | 3 | 69386 | 8.23% | oxidative stress | 1970 (1970) | Cell or Molecular Dysfunction |
| 1970s | 4 | 64508 | 10.5% | logistic regression | 1974 (1974) | Research Activity |
| 1970s | 5 | 60211 | 12.7% | overexpression | 1977 (1977) | Genetic Function |
| 1970s | 6 | 54912 | 14.7% | upregulation | 1979 (1972) | Genetic Function |
| 1970s | 7 | 52661 | 16.6% | metaanalysis | 1977 (1975) | Research Activity |
| 1970s | 8 | 45151 | 18.2% | reactive oxygen species | 1977 (1949) | Element, Ion, or Isotope |
| 1970s | 9 | 37086 | 19.5% | upregulation | 1979 (1979) | Molecular Function |
| 1970s | 10 | 36853 | 20.8% | mrna expression | 1979 (1949) | Genetic Function |
| 1970s | 11 | 35969 | 22.1% | protein expression | 1976 (1949) | Genetic Function |
| 1970s | 12 | 28786 | 23.2% | logistic regression analysis | 1974 (1974) | Research Activity |
| 1970s | 13 | 23896 | 24.0% | overexpress | 1977 (1977) | Genetic Function |
| 1970s | 14 | 23598 | 24.9% | randomized controlled trial | 1970 (1970) | Research Activity |
| 1970s | 15 | 22945 | 25.7% | downregulation | 1977 (1977) | Molecular Function |
| 1970s | 16 | 20421 | 26.5% | CD8 | 1979 (1976) | Immunologic Factor |
| 1970s | 17 | 19721 | 27.2% | T cells | 1970 (1967) | Cell |
| 1970s | 18 | 17182 | 27.8% | FTIR | 1975 (1972) | Research Activity |
| 1970s | 19 | 16625 | 28.4% | ANOVA | 1971 (1971) | Gene or Genome |
| 1970s | 20 | 16541 | 29.0% | ANOVA | 1971 (1971) | Amino Acid, Peptide, or Protein |
| 1960s | 1 | 86187 | 3.94% | mrna | 1964 (1961) | Nucleic Acid, Nucleoside, or Nucleotide |
| 1960s | 2 | 83131 | 7.75% | targeted | 1969 (1969) | Cell Function |
| 1960s | 3 | 60306 | 10.5% | gene expression | 1961 (1949) | Genetic Function |
| 1960s | 4 | 45781 | 12.6% | crosssectional study | 1961 (1954) | Research Activity |
| 1960s | 5 | 32420 | 14.0% | genomic | 1961 (1949) | Gene or Genome |
| 1960s | 6 | 31775 | 15.5% | transcriptional | 1966 (1949) | Genetic Function |
| 1960s | 7 | 24042 | 16.6% | extracellular matrix | 1962 (1952) | Tissue |
| 1960s | 8 | 18772 | 17.5% | transcripts | 1962 (1949) | Nucleic Acid, Nucleoside, or Nucleotide |
| 1960s | 9 | 18060 | 18.3% | casecontrol study | 1967 (1967) | Research Activity |
| 1960s | 10 | 18018 | 19.1% | phylogenetic analysis | 1964 (1949) | Research Activity |
| 1960s | 11 | 16077 | 19.9% | 16S rrna | 1968 (1966) | Nucleic Acid, Nucleoside, or Nucleotide |
| 1960s | 12 | 15664 | 20.6% | retrospective cohort study | 1966 (1966) | Research Activity |
| 1960s | 13 | 15057 | 21.3% | translational | 1963 (1949) | Genetic Function |
| 1960s | 14 | 14371 | 21.9% | chiral | 1968 (1949) | Phenomenon or Process |
| 1960s | 15 | 14303 | 22.6% | COPD | 1967 (1967) | Gene or Genome |
| 1960s | 16 | 14096 | 23.2% | DNA damage | 1965 (1965) | Cell or Molecular Dysfunction |
| 1960s | 17 | 13568 | 23.8% | immunosuppression | 1964 (1964) | Organism Function |
| 1960s | 18 | 13020 | 24.4% | transfection | 1966 (1966) | Molecular Biology Research Technique |
| 1960s | 19 | 11614 | 25.0% | drug discovery | 1964 (1964) | Research Activity |
| 1960s | 20 | 10962 | 25.5% | eukaryotes | 1968 (1956) | Eukaryote |
| 1950s | 1 | 64715 | 3.69% | randomized | 1953 (1949) | Research Activity |
| 1950s | 2 | 59508 | 7.09% | recombinant | 1951 (1951) | Organism |
| 1950s | 3 | 56353 | 10.3% | simulations | 1954 (1949) | Research Activity |
| 1950s | 4 | 33856 | 12.2% | selfreport | 1953 (1953) | Research Activity |
| 1950s | 5 | 33379 | 14.1% | prospective study | 1954 (1954) | Research Activity |
| 1950s | 6 | 25870 | 15.6% | polymorphisms | 1952 (1949) | Genetic Function |
| 1950s | 7 | 17037 | 16.5% | cterminal | 1952 (1952) | Amino Acid Sequence |
| 1950s | 8 | 16458 | 17.5% | oligomer | 1958 (1958) | Amino Acid Sequence |
| 1950s | 9 | 16243 | 18.4% | reperfusion | 1952 (1952) | Biologic Function |
| 1950s | 10 | 14279 | 19.2% | cloned | 1958 (1949) | Cell |
| 1950s | 11 | 14238 | 20.0% | binding sites | 1952 (1949) | Receptor |
| 1950s | 12 | 13733 | 20.8% | ecosystems | 1959 (1956) | Natural Phenomenon or Process |
| 1950s | 13 | 13641 | 21.6% | genetic diversity | 1959 (1949) | Natural Phenomenon or Process |
| 1950s | 14 | 11997 | 22.3% | binding site | 1952 (1952) | Amino Acid Sequence |
| 1950s | 15 | 11940 | 23.0% | genomes | 1957 (1949) | Gene or Genome |
| 1950s | 16 | 11873 | 23.7% | data collection | 1952 (1952) | Research Activity |
| 1950s | 17 | 11186 | 24.3% | hepatocytes | 1956 (1949) | Cell |
| 1950s | 18 | 11107 | 24.9% | binding site | 1952 (1949) | Receptor |
| 1950s | 19 | 10352 | 25.5% | placebocontrolled | 1954 (1953) | Research Activity |
| 1950s | 20 | 10160 | 26.1% | exon | 1950 (1950) | Nucleic Acid, Nucleoside, or Nucleotide |
| | | | | | | |
| | | | | | | |
| MISCELLANEOUS (4th of 4 idea category groups) | | | | | | |
| 2010s | 1 | 1105 | 3.28% | patient protection and affordable care act | 2010 (1981) | Regulation or Law |
| 2010s | 2 | 569 | 4.98% | cha2ds2vasc score | 2010 (2010) | Intellectual Product |
| 2010s | 3 | 262 | 5.76% | HASBLED score | 2011 (2011) | Intellectual Product |
| 2010s | 4 | 173 | 6.27% | PAM50 | 2010 (2010) | Functional Concept |
| 2010s | 5 | 161 | 6.75% | vaping | 2011 (1970) | Individual Behavior |
| 2010s | 6 | 155 | 7.21% | affordable care acts | 2010 (1981) | Regulation or Law |
| 2010s | 7 | 148 | 7.65% | human connectome project | 2011 (2011) | Biomedical Occupation or Discipline |
| 2010s | 8 | 100 | 7.95% | prostate imaging reporting and data system | 2013 (2012) | Classification |
| 2010s | 9 | 100 | 8.25% | prostate imaging reporting and data system | 2013 (2013) | Intellectual Product |
| 2010s | 10 | 79 | 8.48% | PIRADS | 2012 (2012) | Classification |
| 2010s | 11 | 76 | 8.71% | level of evidence II | 2010 (2010) | Conceptual Entity |
| 2010s | 12 | 74 | 8.93% | soft robotics | 2011 (2001) | Occupation or Discipline |
| 2010s | 13 | 73 | 9.15% | 3D printed model | 2013 (2013) | Manufactured Object |
| 2010s | 14 | 69 | 9.35% | operation new dawn | 2011 (2011) | Idea or Concept |
| 2010s | 15 | 58 | 9.53% | activity trackers | 2012 (2012) | Manufactured Object |
| 2010s | 16 | 56 | 9.69% | standard uptake value ratio | 2010 (1991) | Quantitative Concept |
| 2010s | 17 | 51 | 9.84% | national center for advancing translational sciences | 2011 (1990) | Health Care Related Organization |
| 2010s | 18 | 51 | 10.0% | groningen frailty indicator | 2010 (2010) | Intellectual Product |
| 2010s | 19 | 49 | 10.1% | grch37 | 2010 (2010) | Intellectual Product |
| 2010s | 20 | 48 | 10.2% | nannochloropsis oceanica | 2011 (2011) | Plant |
| 2000s | 1 | 8547 | 5.76% | regenerative medicine | 2000 (2000) | Biomedical Occupation or Discipline |
| 2000s | 2 | 7193 | 10.6% | metabolomics | 2000 (1951) | Biomedical Occupation or Discipline |
| 2000s | 3 | 6827 | 15.2% | gene ontology | 2000 (2000) | Intellectual Product |
| 2000s | 4 | 3136 | 17.3% | metagenomic | 2000 (1987) | Occupation or Discipline |
| 2000s | 5 | 2883 | 19.2% | metabolomic | 2000 (1951) | Biomedical Occupation or Discipline |
| 2000s | 6 | 2686 | 21.1% | DSM5 | 2000 (2000) | Intellectual Product |
| 2000s | 7 | 2172 | 22.5% | smartphone | 2004 (2004) | Manufactured Object |
| 2000s | 8 | 1791 | 23.7% | metagenomics | 2003 (1987) | Occupation or Discipline |
| 2000s | 9 | 1525 | 24.8% | theranostic | 2000 (2000) | Biomedical Occupation or Discipline |
| 2000s | 10 | 1503 | 25.8% | smartphones | 2004 (2004) | Manufactured Object |
| 2000s | 11 | 1438 | 26.7% | MELD score | 2001 (2001) | Intellectual Product |
| 2000s | 12 | 1401 | 27.7% | RECIST | 2000 (2000) | Intellectual Product |
| 2000s | 13 | 1287 | 28.6% | nanoribbons | 2000 (2000) | Manufactured Object |
| 2000s | 14 | 1283 | 29.4% | model for endstage liver disease | 2001 (1988) | Classification |
| 2000s | 15 | 1200 | 30.2% | response evaluation criteria in solid tumors | 2000 (2000) | Intellectual Product |
| 2000s | 16 | 1171 | 31.0% | hapmap | 2002 (2002) | Organism Attribute |
| 2000s | 17 | 1100 | 31.8% | common terminology criteria for adverse events | 2003 (1991) | Intellectual Product |
| 2000s | 18 | 1042 | 32.5% | centers for medicare and medicaid services | 2001 (1977) | Health Care Related Organization |
| 2000s | 19 | 1032 | 33.2% | montreal cognitive assessment | 2005 (1960) | Intellectual Product |
| 2000s | 20 | 939 | 33.8% | agency for healthcare research and quality | 2000 (1990) | Health Care Related Organization |
| 1990s | 1 | 27976 | 6.07% | microarray | 1992 (1992) | Manufactured Object |
| 1990s | 2 | 16380 | 9.63% | proteomics | 1997 (1997) | Biomedical Occupation or Discipline |
| 1990s | 3 | 15358 | 12.9% | proteomic | 1997 (1997) | Biomedical Occupation or Discipline |
| 1990s | 4 | 11317 | 15.4% | knockout mice | 1992 (1978) | Mammal |
| 1990s | 5 | 9075 | 17.4% | nanomaterials | 1994 (1986) | Manufactured Object |
| 1990s | 6 | 6885 | 18.9% | nanowires | 1993 (1993) | Manufactured Object |
| 1990s | 7 | 6540 | 20.3% | SF36 | 1991 (1991) | Intellectual Product |
| 1990s | 8 | 5950 | 21.6% | nanotechnology | 1991 (1991) | Occupation or Discipline |
| 1990s | 9 | 5366 | 22.7% | innate immune response | 1993 (1949) | Organism Attribute |
| 1990s | 10 | 5090 | 23.8% | nanorods | 1999 (1999) | Manufactured Object |
| 1990s | 11 | 4868 | 24.9% | men who have sex with men | 1991 (1991) | Population Group |
| 1990s | 12 | 4702 | 25.9% | systems biology | 1993 (1993) | Biomedical Occupation or Discipline |
| 1990s | 13 | 4546 | 26.9% | evidencebased practice | 1993 (1993) | Functional Concept |
| 1990s | 14 | 4522 | 27.9% | support vector machine | 1998 (1998) | Quantitative Concept |
| 1990s | 15 | 4458 | 28.9% | centers for disease control and prevention | 1991 (1971) | Health Care Related Organization |
| 1990s | 16 | 4438 | 29.8% | nanofibers | 1994 (1994) | Manufactured Object |
| 1990s | 17 | 4331 | 30.8% | clinical practice guidelines | 1990 (1990) | Intellectual Product |
| 1990s | 18 | 4168 | 31.7% | evidencebased medicine | 1992 (1992) | Biomedical Occupation or Discipline |
| 1990s | 19 | 3897 | 32.5% | affymetrix | 1995 (1995) | Health Care Related Organization |
| 1990s | 20 | 3826 | 33.3% | innate immune responses | 1995 (1949) | Organism Attribute |
| 1980s | 1 | 39848 | 4.54% | hazard ratio | 1980 (1980) | Quantitative Concept |
| 1980s | 2 | 39599 | 9.06% | comorbidities | 1986 (1970) | Idea or Concept |
| 1980s | 3 | 17746 | 11.0% | progressionfree survival | 1983 (1983) | Quantitative Concept |
| 1980s | 4 | 15786 | 12.8% | stakeholder | 1981 (1981) | Conceptual Entity |
| 1980s | 5 | 15321 | 14.6% | bioinformatics | 1989 (1988) | Biomedical Occupation or Discipline |
| 1980s | 6 | 15038 | 16.3% | healthrelated quality of life | 1982 (1982) | Idea or Concept |
| 1980s | 7 | 13421 | 17.8% | molecular dynamics simulations | 1981 (1973) | Machine Activity |
| 1980s | 8 | 11555 | 19.2% | focus groups | 1980 (1977) | Group |
| 1980s | 9 | 10754 | 20.4% | biodiversity | 1988 (1968) | Qualitative Concept |
| 1980s | 10 | 10658 | 21.6% | transgenic mice | 1982 (1982) | Mammal |
| 1980s | 11 | 10558 | 22.8% | electronic database | 1980 (1980) | Intellectual Product |
| 1980s | 12 | 8609 | 23.8% | microfluidic | 1988 (1988) | Occupation or Discipline |
| 1980s | 13 | 8220 | 24.7% | nanostructures | 1986 (1986) | Manufactured Object |
| 1980s | 14 | 6864 | 25.5% | bioinformatic | 1988 (1988) | Biomedical Occupation or Discipline |
| 1980s | 15 | 6864 | 26.3% | primary outcome measure | 1981 (1981) | Qualitative Concept |
| 1980s | 16 | 6852 | 27.1% | propensity score | 1987 (1987) | Quantitative Concept |
| 1980s | 17 | 6804 | 27.8% | gene expression profiles | 1989 (1989) | Quantitative Concept |
| 1980s | 18 | 6312 | 28.6% | nanostructure | 1986 (1986) | Manufactured Object |
| 1980s | 19 | 6123 | 29.3% | quantum dots | 1987 (1987) | Manufactured Object |
| 1980s | 20 | 6114 | 30.0% | african americans | 1980 (1949) | Population Group |
| 1970s | 1 | 109968 | 4.43% | targeting | 1971 (1949) | Functional Concept |
| 1970s | 2 | 71641 | 7.32% | odds ratio | 1970 (1970) | Quantitative Concept |
| 1970s | 3 | 62961 | 9.86% | magnetic resonance imaging | 1978 (1978) | Professional or Occupational Group |
| 1970s | 4 | 58091 | 12.2% | expression level | 1979 (1979) | Quantitative Concept |
| 1970s | 5 | 54297 | 14.4% | metaanalysis | 1977 (1977) | Intellectual Product |
| 1970s | 6 | 52768 | 16.5% | nanoparticles | 1978 (1978) | Manufactured Object |
| 1970s | 7 | 34573 | 17.9% | inclusion criteria | 1976 (1949) | Qualitative Concept |
| 1970s | 8 | 33562 | 19.2% | dataset | 1970 (1949) | Intellectual Product |
| 1970s | 9 | 32970 | 20.6% | apoptotic | 1972 (1972) | Qualitative Concept |
| 1970s | 10 | 26379 | 21.6% | scenarios | 1974 (1949) | Functional Concept |
| 1970s | 11 | 25440 | 22.7% | IC50 | 1970 (1965) | Quantitative Concept |
| 1970s | 12 | 24189 | 23.6% | gold standard | 1979 (1979) | Qualitative Concept |
| 1970s | 13 | 23802 | 24.6% | databases | 1971 (1949) | Intellectual Product |
| 1970s | 14 | 22336 | 25.5% | transgenic | 1972 (1972) | Animal |
| 1970s | 15 | 22291 | 26.4% | odds ratios | 1978 (1970) | Quantitative Concept |
| 1970s | 16 | 18359 | 27.1% | comorbidity | 1970 (1970) | Idea or Concept |
| 1970s | 17 | 18025 | 27.9% | patient outcome | 1970 (1970) | Idea or Concept |
| 1970s | 18 | 17325 | 28.6% | risk assessment | 1973 (1973) | Intellectual Product |
| 1970s | 19 | 17232 | 29.3% | scaffolds | 1975 (1949) | Manufactured Object |
| 1970s | 20 | 16241 | 29.9% | nonsmall cell lung cancer | 1976 (1976) | Conceptual Entity |
| 1960s | 1 | 89643 | 3.02% | targeted | 1969 (1949) | Functional Concept |
| 1960s | 2 | 56711 | 4.93% | software | 1965 (1960) | Manufactured Object |
| 1960s | 3 | 54652 | 6.77% | ongoing | 1960 (1949) | Idea or Concept |
| 1960s | 4 | 53347 | 8.57% | genomic | 1961 (1961) | Biomedical Occupation or Discipline |
| 1960s | 5 | 51686 | 10.3% | optimization | 1960 (1960) | Activity |
| 1960s | 6 | 51434 | 12.0% | sequencing | 1962 (1962) | Functional Concept |
| 1960s | 7 | 44072 | 13.5% | sequencing | 1962 (1962) | Intellectual Product |
| 1960s | 8 | 38380 | 14.8% | time point | 1960 (1955) | Temporal Concept |
| 1960s | 9 | 36566 | 16.0% | dosedependent | 1960 (1960) | Quantitative Concept |
| 1960s | 10 | 35745 | 17.2% | animal models | 1962 (1954) | Animal |
| 1960s | 11 | 32058 | 18.3% | colorectal cancer | 1962 (1962) | Conceptual Entity |
| 1960s | 12 | 30573 | 19.3% | automated | 1960 (1949) | Functional Concept |
| 1960s | 13 | 28157 | 20.3% | transcripts | 1962 (1962) | Intellectual Product |
| 1960s | 14 | 27647 | 21.2% | providers | 1960 (1949) | Functional Concept |
| 1960s | 15 | 26162 | 22.1% | colorectal cancer | 1962 (1962) | Intellectual Product |
| 1960s | 16 | 24749 | 22.9% | overall survival | 1963 (1963) | Quantitative Concept |
| 1960s | 17 | 23419 | 23.7% | ethnicity | 1966 (1966) | Qualitative Concept |
| 1960s | 18 | 22479 | 24.5% | algorithms | 1963 (1949) | Intellectual Product |
| 1960s | 19 | 21854 | 25.2% | ex vivo | 1964 (1964) | Functional Concept |
| 1960s | 20 | 19862 | 25.9% | ethnicity | 1966 (1949) | Population Group |
| 1950s | 1 | 101395 | 2.49% | risk factors | 1959 (1959) | Intellectual Product |
| 1950s | 2 | 67299 | 4.15% | quality of life | 1959 (1959) | Idea or Concept |
| 1950s | 3 | 62097 | 5.68% | encoding | 1956 (1953) | Activity |
| 1950s | 4 | 62041 | 7.21% | encoding | 1956 (1956) | Idea or Concept |
| 1950s | 5 | 52107 | 8.50% | downstream | 1950 (1950) | Spatial Concept |
| 1950s | 6 | 50147 | 9.73% | researchers | 1954 (1949) | Professional or Occupational Group |
| 1950s | 7 | 46112 | 10.8% | documented | 1950 (1950) | Intellectual Product |
| 1950s | 8 | 41331 | 11.8% | technologies | 1956 (1949) | Occupation or Discipline |
| 1950s | 9 | 38756 | 12.8% | options | 1950 (1949) | Functional Concept |
| 1950s | 10 | 33842 | 13.6% | modulating | 1955 (1949) | Spatial Concept |
| 1950s | 11 | 31872 | 14.4% | intraoperative | 1950 (1950) | Temporal Concept |
| 1950s | 12 | 30024 | 15.2% | categorized | 1957 (1952) | Activity |
| 1950s | 13 | 29996 | 15.9% | encode | 1953 (1953) | Activity |
| 1950s | 14 | 29951 | 16.6% | older adult | 1951 (1951) | Age Group |
| 1950s | 15 | 28595 | 17.3% | lifestyle | 1959 (1956) | Social Behavior |
| 1950s | 16 | 28278 | 18.0% | perioperative | 1957 (1957) | Temporal Concept |
| 1950s | 17 | 26289 | 18.7% | older adult | 1951 (1949) | Population Group |
| 1950s | 18 | 25647 | 19.3% | emergency department | 1957 (1957) | Health Care Related Organization |
| 1950s | 19 | 25130 | 19.9% | encoded | 1953 (1953) | Activity |
| 1950s | 20 | 25019 | 20.6% | multidisciplinary | 1952 (1949) | Occupational Activity |

**Table S4.** Bootstrapped Confidence Intervals and Robustness of Overall Frontier Positions of Nations to Alternative Specifications.

| (1a) | (1b) | (1c) | (1d) | | (2) | (3) | (4) | (5) | (6) | (7) | (8) | (9) |
|---|---|---|---|---|---|---|---|---|---|---|---|---|
| Location | Number of Contributions | 2010s; same as column 1c in Table 1 | Bootstrapped 95% Confidence Intervals | | Set missing values equal to 0 | Weight by number of own contri-butions | Use UMLS synonym data to determine cohort of each term | Top 20% novel status used | Top 10% novel status used | Top 1% novel status used | All papers (not just original res. papers) | No upper, lower limits on number of char-acters |
| UNITED STATES | 2853661 | 108 | (106..109) | | 108 | 107 | 102 | 105 | 114 | 108 | 107 | 108 |
| SOUTH KOREA | 374227 | 107 | (104..110) | | 105 | 107 | 108 | 108 | 99 | 107 | 108 | 106 |
| SINGAPORE | 52541 | 105 | (100..111) | | 98 | 108 | 99 | 103 | 118 | 106 | 106 | 105 |
| TAIWAN | 177229 | 104 | (101..107) | | 102 | 103 | 107 | 107 | 98 | 103 | 105 | 104 |
| IRELAND | 39495 | 103 | (97..108) | | 95 | 100 | 98 | 98 | 109 | 101 | 102 | 100 |
| BELGIUM | 95644 | 102 | (99..106) | | 99 | 102 | 99 | 100 | 104 | 104 | 101 | 101 |
| ITALY | 384029 | 102 | (99..104) | | 101 | 103 | 100 | 102 | 100 | 102 | 103 | 102 |
| CHINA | 1734035 | 101 | (99..103) | | 101 | 102 | 105 | 102 | 95 | 102 | 101 | 101 |
| CANADA | 375846 | 101 | (99..104) | | 101 | 101 | 98 | 100 | 105 | 99 | 101 | 102 |
| JAPAN | 554589 | 100 | (98..103) | | 99 | 103 | 100 | 100 | 100 | 100 | 101 | 100 |
| UNITED KINGDOM | 494917 | 100 | (98..103) | | 100 | 100 | 97 | 98 | 105 | 100 | 100 | 100 |
| NETHERLANDS | 233631 | 100 | (97..103) | | 99 | 100 | 99 | 99 | 100 | 100 | 100 | 100 |
| GERMANY | 539888 | 100 | (98..102) | | 99 | 99 | 97 | 98 | 101 | 100 | 99 | 99 |
| SWITZERLAND | 123779 | 100 | (96..103) | | 97 | 100 | 97 | 98 | 102 | 100 | 101 | 99 |
| SAUDI ARABIA | 34855 | 99 | (94..106) | | 90 | 94 | 95 | 95 | 82 | 97 | 98 | 98 |
| FINLAND | 59534 | 99 | (94..103) | | 94 | 98 | 100 | 98 | 96 | 98 | 99 | 98 |
| NORWAY | 63699 | 98 | (93..103) | | 94 | 95 | 96 | 95 | 96 | 97 | 97 | 99 |
| SOUTH AFRICA | 43179 | 98 | (92..104) | | 90 | 101 | 99 | 103 | 92 | 100 | 96 | 97 |
| SPAIN | 278504 | 98 | (95..100) | | 97 | 97 | 98 | 98 | 92 | 99 | 97 | 97 |
| CZECH REPUBLIC | 44024 | 97 | (92..102) | | 89 | 102 | 98 | 99 | 92 | 98 | 99 | 97 |
| AUSTRALIA | 320955 | 97 | (95..99) | | 96 | 97 | 97 | 97 | 96 | 98 | 97 | 97 |
| SWEDEN | 138949 | 96 | (92..99) | | 94 | 96 | 96 | 95 | 92 | 96 | 96 | 96 |
| AUSTRIA | 65039 | 96 | (91..100) | | 91 | 97 | 96 | 97 | 99 | 96 | 96 | 96 |
| DENMARK | 105066 | 95 | (92..99) | | 93 | 95 | 97 | 96 | 99 | 97 | 95 | 95 |
| FRANCE | 305065 | 95 | (93..98) | | 94 | 96 | 95 | 96 | 98 | 98 | 95 | 96 |
| POLAND | 113074 | 93 | (89..96) | | 90 | 93 | 94 | 93 | 90 | 90 | 93 | 93 |
| THAILAND | 40080 | 93 | (87..98) | | 85 | 92 | 96 | 95 | 84 | 93 | 95 | 92 |
| HUNGARY | 28574 | 92 | (86..99) | | 82 | 94 | 94 | 93 | 86 | 95 | 94 | 90 |
| ISRAEL | 76781 | 92 | (88..96) | | 89 | 93 | 95 | 94 | 90 | 94 | 92 | 92 |
| OTHER EUROPE | 107712 | 90 | (87..94) | | 88 | 90 | 92 | 90 | 88 | 90 | 92 | 90 |
| NEW ZEALAND | 38946 | 90 | (84..96) | | 83 | 91 | 90 | 90 | 96 | 93 | 98 | 93 |
| TURKEY | 157825 | 90 | (86..94) | | 87 | 91 | 95 | 94 | 84 | 90 | 90 | 89 |
| RUSSIA | 51759 | 89 | (83..95) | | 79 | 89 | 87 | 86 | 90 | 86 | 90 | 89 |
| CHILE | 23794 | 89 | (82..97) | | 78 | 88 | 95 | 92 | 71 | 96 | 87 | 89 |
| GREECE | 46646 | 89 | (84..93) | | 83 | 92 | 98 | 93 | 81 | 92 | 90 | 90 |
| MALAYSIA | 37997 | 87 | (82..93) | | 79 | 89 | 91 | 86 | 86 | 84 | 92 | 87 |
| PORTUGAL | 65523 | 86 | (82..91) | | 82 | 86 | 91 | 89 | 87 | 86 | 88 | 87 |
| OTHER ASIA | 60973 | 86 | (81..90) | | 81 | 83 | 91 | 87 | 82 | 87 | 86 | 85 |
| INDIA | 291215 | 83 | (80..86) | | 81 | 81 | 89 | 86 | 80 | 84 | 84 | 82 |
| BRAZIL | 274896 | 83 | (80..85) | | 81 | 85 | 91 | 87 | 76 | 79 | 84 | 83 |
| PAKISTAN | 27511 | 83 | (75..91) | | 69 | 81 | 85 | 83 | 83 | 81 | 81 | 83 |
| MEXICO | 54997 | 81 | (77..86) | | 76 | 83 | 89 | 85 | 75 | 79 | 82 | 80 |
| IRAN | 121035 | 78 | (74..82) | | 76 | 79 | 89 | 84 | 70 | 79 | 79 | 78 |
| OTHER AMERICAS | 30787 | 77 | (71..83) | | 70 | 86 | 88 | 82 | 78 | 73 | 81 | 77 |
| ARGENTINA | 40775 | 77 | (72..82) | | 70 | 79 | 84 | 80 | 77 | 72 | 81 | 78 |
| EGYPT | 48649 | 75 | (69..80) | | 68 | 74 | 89 | 83 | 65 | 78 | 75 | 74 |
| OTHER AFRICA | 90041 | 70 | (65..74) | | 65 | 72 | 81 | 76 | 69 | 70 | 71 | 70 |

Notes to Table S4:

All numbers are calculated based on papers published during 2015-2016.

Column 1a: Location.

Column 1b: Number of contributions based on which the edge factor in column (1c) is calculated. See notes to Table 1.

Column 1c: Edge factor for the baseline specification.

Column 1d: Bootstrapped 95% confidence interval for the edge factor in the baseline specification.

Column 2: When there are no observations for an (idea category, research area) pair for a location, the edge factor for that that (idea category, research area) pair is set to 0; in the baseline specification (shown in column 1c) the edge factor is set to the weighted average of the edge factor for all other (idea category, research area) pairs.

Column 3: When the overall edge factor is calculated for a location, the weight of the edge factor for each (idea category, research area) pair is the location's own number of papers linked to that (idea category, research area) pair; in the baseline specification (shown in column 1c) the weight is the number of papers from any location that are linked to that (idea category, research area) pair.

Column 4: The vintage of each UMLS term is determined based on the earliest year of appearance of the UMLS term or any of its synonyms (as indicated in the UMLS); in the baseline specification (shown in column 1c) vintage is determined based on the earliest year of appearance of the UMLS term.

Column 5: When the dummy variable that indicates the novelty of a contribution relative to other contributions in the comparison group is constructed, a 20% cutoff level is used, so that the 20% of the contributions with the most recent cohort are assigned the novel status; in the baseline specification (shown in column 1c) the corresponding cutoff is 5%.

Column 6: Same as Column (5) but now a 10% cutoff is used.

Column 7: Same as Column (5) but now a 1% cutoff is used.

Column 8: The analysis includes all types of publications in MEDLINE; in the baseline specification (shown in column 1c) only original research papers are considered.

Column 9: The analysis includes also those papers for which MEDLINE has either less than 200 characters of text or more than 5000 characters of text; in the baseline specification (shown in column 1c) only those original research papers are included for which the text information in MEDLINE falls within those bounds.

**Table S5**. Overall Scientific Frontier Positions of Nations by Time Period.

| (1a) | (1b) | (1c) | | (2a) | (2b) | (2c) | (2d) | (2e) | (2f) |
|---|---|---|---|---|---|---|---|---|---|
| Location | Number of Contributions | 2015-6 | | 1990-94 with 1990s weights | 1995-9, with 1990s weights | 2000-4, with 1990s weights | 2005-9, with 1990s weights | 2010-4, with 1990s weights | 2015-6, with 1990s weights |
| UNITED STATES | 2853661 | 108 | | 107 | 108 | 109 | 109 | 109 | 109 |
| SOUTH KOREA | 374227 | 107 | | 82 | 87 | 102 | 106 | 105 | 107 |
| SINGAPORE | 52541 | 105 | | 89 | 105 | 102 | 108 | 109 | 104 |
| TAIWAN | 177229 | 104 | | 90 | 85 | 89 | 95 | 101 | 105 |
| IRELAND | 39495 | 103 | | 74 | 84 | 97 | 100 | 106 | 99 |
| BELGIUM | 95644 | 102 | | 101 | 101 | 104 | 108 | 107 | 106 |
| ITALY | 384029 | 102 | | 94 | 95 | 95 | 99 | 102 | 103 |
| CHINA | 1734035 | 101 | | 79 | 82 | 98 | 97 | 98 | 101 |
| CANADA | 375846 | 101 | | 93 | 96 | 98 | 100 | 101 | 100 |
| JAPAN | 554589 | 100 | | 95 | 96 | 100 | 101 | 101 | 102 |
| UNITED KINGDOM | 494917 | 100 | | 103 | 104 | 103 | 103 | 104 | 101 |
| NETHERLANDS | 233631 | 100 | | 93 | 94 | 99 | 103 | 102 | 100 |
| GERMANY | 539888 | 100 | | 96 | 97 | 101 | 102 | 102 | 103 |
| SWITZERLAND | 123779 | 100 | | 105 | 107 | 105 | 104 | 104 | 103 |
| SAUDI ARABIA | 34855 | 99 | | 83 | 75 | 65 | 85 | 87 | 95 |
| FINLAND | 59534 | 99 | | 95 | 95 | 96 | 93 | 96 | 99 |
| NORWAY | 63699 | 98 | | 90 | 92 | 90 | 97 | 99 | 100 |
| SOUTH AFRICA | 43179 | 98 | | 75 | 76 | 84 | 84 | 84 | 96 |
| SPAIN | 278504 | 98 | | 79 | 85 | 90 | 92 | 96 | 99 |
| CZECH REPUBLIC | 44024 | 97 | | 70 | 75 | 87 | 90 | 94 | 97 |
| AUSTRALIA | 320955 | 97 | | 92 | 94 | 96 | 96 | 98 | 98 |
| SWEDEN | 138949 | 96 | | 87 | 87 | 90 | 94 | 97 | 98 |
| AUSTRIA | 65039 | 96 | | 99 | 98 | 99 | 102 | 99 | 99 |
| DENMARK | 105066 | 95 | | 89 | 89 | 91 | 94 | 97 | 96 |
| FRANCE | 305065 | 95 | | 97 | 96 | 97 | 101 | 99 | 97 |
| POLAND | 113074 | 93 | | 59 | 68 | 73 | 77 | 87 | 92 |
| THAILAND | 40080 | 93 | | 79 | 73 | 73 | 75 | 79 | 85 |
| HUNGARY | 28574 | 92 | | 72 | 75 | 80 | 83 | 85 | 94 |
| ISRAEL | 76781 | 92 | | 79 | 78 | 87 | 93 | 94 | 95 |
| OTHER EUROPE | 107712 | 90 | | 74 | 73 | 76 | 77 | 82 | 89 |
| NEW ZEALAND | 38946 | 90 | | 92 | 92 | 92 | 93 | 95 | 87 |
| TURKEY | 157825 | 90 | | 70 | 67 | 72 | 69 | 82 | 86 |
| RUSSIA | 51759 | 89 | | 57 | 68 | 68 | 65 | 81 | 89 |
| CHILE | 23794 | 89 | | 49 | 62 | 74 | 74 | 85 | 89 |
| GREECE | 46646 | 89 | | 71 | 78 | 78 | 80 | 84 | 90 |
| MALAYSIA | 37997 | 87 | | 57 | 77 | 94 | 68 | 80 | 81 |
| PORTUGAL | 65523 | 86 | | 81 | 82 | 74 | 88 | 93 | 84 |
| OTHER ASIA | 60973 | 86 | | 70 | 64 | 70 | 69 | 76 | 85 |
| INDIA | 291215 | 83 | | 57 | 61 | 65 | 68 | 74 | 78 |
| BRAZIL | 274896 | 83 | | 81 | 74 | 74 | 75 | 77 | 79 |
| PAKISTAN | 27511 | 83 | | 53 | 56 | 57 | 75 | 77 | 81 |
| MEXICO | 54997 | 81 | | 73 | 69 | 70 | 73 | 74 | 79 |
| IRAN | 121035 | 78 | | 32 | 36 | 56 | 66 | 72 | 74 |
| OTHER AMERICAS | 30787 | 77 | | 81 | 84 | 71 | 77 | 74 | 70 |
| ARGENTINA | 40775 | 77 | | 58 | 65 | 69 | 69 | 75 | 74 |
| EGYPT | 48649 | 75 | | 49 | 67 | 62 | 65 | 67 | 71 |
| OTHER AFRICA | 90041 | 70 | | 77 | 76 | 66 | 67 | 67 | 63 |

Notes to Table S5:

Column 1a: Location.

Column 1b: Number of contributions based on which the edge factor in column (1c) is calculated. See notes to Table 1.

Weights below refer to how the edge factor for each (idea category, research area) pair is weighted when the overall edge factor for a location is calculated. When “2015-6 weights” are used, the weight for each (idea category, research area) pair is the total number of papers published during 2015-2016 that are linked to that (idea category, research area) pair.

Column 1c: Edge factors for 2015-2016 (calculated using 2015-6 weights).

Column 2a. Edge factors for 1998-1994, calculated using 1990s weights.

Column 2b. Edge factors for 1995-1999, calculated using 1990s weights.

Column 2c. Edge factors for 2000-2004, calculated using 1990s weights.

Column 2d. Edge factors for 2005-2009, calculated using 1990s weights.

Column 2e. Edge factors for 2010-2014, calculated using 1990s weights.

Column 2f. Edge factors for 2015-2016, calculated using 1990s weights.